\documentclass[aps,prd,preprintnumbers,showpacs,superscriptaddress,nofootinbib,amsmath,amssymb,floats,floatfix,showkeys,notitlepage,longbibliography]{revtex4-1}
\usepackage[table]{xcolor} 
\usepackage{array} 
\renewcommand{\arraystretch}{1.5} 
\definecolor{lightblue}{rgb}{0.62, 0.83, 0.96} 
\definecolor{lightgray}{rgb}{0.93, 0.97, 0.99} 
\usepackage{comment}
\usepackage{graphicx}
\usepackage{subfigure}
\usepackage{palatino}
\usepackage[commandnameprefix=always]{changes}
\usepackage{hyperref}
\hypersetup{colorlinks=true,linkcolor=red,urlcolor=red,citecolor=red}
\usepackage[toc,page]{appendix}
\usepackage[normalem]{ulem}
\usepackage{lipsum}
\usepackage{graphicx}
\usepackage{subfigure}
\usepackage{palatino}
\usepackage{float}
\usepackage{sans}
\usepackage{adjustbox}
\usepackage{latexsym}
\usepackage{amsmath}
\usepackage{amssymb}
\usepackage{amsfonts}
\usepackage{dcolumn}
\usepackage{bm}
\usepackage{tikz}
\usepackage{bigints}
\usepackage{array,tabularx,multirow,booktabs}
\usepackage[tracking=true]{microtype}
\SetTracking{}{500}
\SetTracking{encoding={*}, shape=sc}{40}
\allowdisplaybreaks
\usepackage{adjustbox}
\usepackage{latexsym}
\usepackage{amsmath}
\usepackage{amssymb}
\usepackage{amsfonts}
\usepackage{dcolumn}
\usepackage{bm}
\usepackage{tikz}
\usepackage{bigints}
\usepackage{array,tabularx,multirow,booktabs}
\usepackage[tracking=true]{microtype}
\usepackage{color}
\allowdisplaybreaks
\begin{document}

\title{Testing the Weak Gravity Conjecture via Gravitational Lensing, Black Hole Shadows, and Barrow Thermodynamics in F(R)--Euler--Heisenberg (A)dS Black Holes}

\author{Saeed Noori Gashti}
\email{sn.gashti@du.ac.ir}
\affiliation{School of Physics, Damghan University, Damghan 3671645667, Iran}

\author{\.{I}zzet Sakall{\i}}
\email{izzet.sakalli@emu.edu.tr}
\affiliation{ Physics Department, Eastern Mediterranean University, Famagusta 99628, North Cyprus via Mersin 10, T\"{u}rkiye}

\author{Erdem Sucu}
\email{erdemsc07@gmail.com}
\affiliation{Physics Department, Eastern Mediterranean University,
Famagusta, 99628 North Cyprus, via Mersin 10, T\"{u}rkiye}

\author{Mohammad Reza Alipour}
\email{mohamad.alipour.1994@gmail.com}
\affiliation{School of Physics, Damghan University, Damghan 3671645667, Iran}
\affiliation{Department of Physics, Faculty of Basic
Sciences, University of Mazandaran\\ P. O. Box 47416-95447, Babolsar, Iran}

\author{Ankit Anand}
\email{Anand@iitk.ac.in}
\affiliation{ Department of Physics, Indian Institute of Technology Kanpur, Kanpur 208016, India.}

\author{Mohammad Ali S Afshar}
\email{m.a.s.afshar@gmail.com}
\affiliation{Department of Physics, Faculty of Basic
Sciences, University of Mazandaran\\ P. O. Box 47416-95447, Babolsar, Iran}
\affiliation{School of Physics, Damghan University, Damghan 3671645667, Iran}
\affiliation{Canadian Quantum Research Center, 204-3002 32 Ave Vernon, BC V1T 2L7, Canada}

\author{Behnam Pourhassan}
\email{b.pourhassan@du.ac.ir}
\affiliation{School of Physics, Damghan University, Damghan 3671645667, Iran}
\affiliation{Center for Theoretical Physics, Khazar University, 41 Mehseti Street, Baku, AZ1096, Azerbaijan}

\author{Jafar Sadeghi}
\email{ pouriya@ipm.ir}
\affiliation{Department of Physics, Faculty of Basic
Sciences, University of Mazandaran\\ P. O. Box 47416-95447, Babolsar, Iran}
\affiliation{Canadian Quantum Research Center, 204-3002 32 Ave Vernon, BC V1T 2L7, Canada}
\begin{abstract}
We investigate the interplay of the Weak Gravity Conjecture (WGC) and the Weak Cosmic Censorship Conjecture (WCCC) in $F(R)$--Euler--Heisenberg black holes in Anti--de Sitter and de Sitter backgrounds. The solution is characterized by the electric charge $q$, the $F(R)$ deviation $f_{R_0}$, the Euler--Heisenberg coupling $\lambda$, and the constant scalar curvature $R_0$. We establish a universal entropy--extremality relation that provides thermodynamic evidence for the WGC independently of $f_{R_0}$ and $R_0$. Photon sphere analysis from both geodesic and topological perspectives confirms the simultaneous compatibility of the WGC and WCCC, with the Euler--Heisenberg coupling restoring photon spheres in the naked singularity regime. Gravitational lensing in the strong- and weak-deflection limits reveals that the photon sphere radius is independent of the cosmological background while the critical impact parameter nearly doubles in de Sitter. Black hole shadow images under isotropic accretion are constructed. Within the Barrow entropy framework, we uncover van der Waals--type phase transitions and analyze Joule--Thomson expansion, identifying the small black hole phase as the WGC-compatible thermodynamic regime accessible via isenthalpic cooling.
\end{abstract}

\date{\today}

\keywords{Weak Gravity Conjecture; $F(R)$--Euler--Heisenberg black holes; gravitational lensing and black hole shadows; Barrow entropy phase transitions; photon sphere}

\pacs{}

\maketitle
\tableofcontents
{\color{black}
\section{Introduction}\label{isec1}

The construction of a self-consistent quantum theory of gravity remains one of the deepest open problems in fundamental physics~\cite{01,02}. Despite notable progress in string theory, loop quantum gravity, asymptotic safety, and causal set approaches~\cite{03,04,05}, no experimentally verified framework currently exists. Within this landscape, the swampland program~\cite{1,2,3,4,5,6,7,8} has gained particular prominence as a means of constraining low-energy effective field theories (EFTs) that may descend from a consistent ultraviolet (UV) completion incorporating gravity. The central premise is that internal consistency at low energies does not guarantee compatibility with quantum gravity: many EFTs that appear viable from a bottom-up perspective cannot be embedded into any UV-complete gravitational theory such as string theory~\cite{1,2}. Those EFTs admitting such an embedding constitute the \emph{landscape}, while the remainder---termed the \emph{swampland}---must be discarded. A growing collection of conjectural criteria, drawing support from black hole (BH) thermodynamics, the Anti-de Sitter/Conformal Field Theory (AdS/CFT) correspondence, and explicit string compactifications, has been proposed to separate these two domains~\cite{2,3,4,5,6,7,8}, with far-reaching consequences for cosmology, dark energy, and particle physics~\cite{a,b,f,g,h,i,j,k,l,m,n,o,p,q,r,s,t,u,v,w,x,y,z,aa,bb,cc,dd,ee,ff,gg,hh,ii,jj,kk,ll,mm,nn,oo,pp,qq,rr,ss,tt,uu,vv,ww,xx,yy,zz,aaa,bbb,ccc,ddd,eee,fff,ggg,hhh,iii,jjj,kkk,lll,rrr,sss,ttt,uuu,vvv,www,xxx,zzz,zzzz}.

Among the most influential of these criteria is the WGC~\cite{9,10}, which posits that any consistent quantum gravity theory containing a $U(1)$ gauge symmetry must admit at least one state satisfying $\widehat{q}/m \geq 1$ in appropriate units. Applied to charged BHs, this condition guarantees the existence of decay channels that prevent extremal BHs from becoming absolutely stable remnants---a scenario incompatible with BH thermodynamics and holography~\cite{9,10}. Conceptually, the WGC encodes the principle that gravity must be the weakest force, and it has found diverse applications in axion physics, inflationary model building, dark matter phenomenology, and discussions of the cosmological constant~\cite{9,10}.

A complementary classical consistency condition is the WCCC, due to Penrose~\cite{9}, which asserts that singularities formed through gravitational collapse are generically hidden behind event horizons (EHs). The WCCC safeguards the predictability of general relativity (GR) by preventing naked singularities from forming, and has been tested extensively through thought experiments involving BH accretion and charged particle absorption~\cite{10000,11,12}.

An apparent tension arises when the WGC and WCCC are simultaneously applied to charged BHs, particularly in the Reissner--Nordstr\"{o}m (RN) geometry~\cite{jjj,kkk}. There, an EH exists only when $q/M \leq 1$; exceeding this bound produces a naked singularity, violating the WCCC. Yet the WGC demands particles with $\widehat{q}/m > 1$, which, if absorbed by a nearly extremal BH, could push it into a super-extremal regime ($q > M$), seemingly destroying the horizon. This clash has stimulated considerable recent work~\cite{45'm,45mmm,45m}, showing that additional matter fields, cosmological constants, and higher-curvature corrections can modify the extremality bound and reconcile the two conjectures.

A thermodynamic formulation of this interplay was developed by Goon and Penco~\cite{Goon:2019faz}, who showed that quantum corrections to extremal BHs induce a universal relation: the extremal mass shift is proportional to the entropy correction with a \emph{negative} coefficient. This implies that quantum effects simultaneously lower the extremality bound and increase the entropy, destabilizing extremal configurations in agreement with the WGC. This pattern has since been confirmed across a broad class of asymptotically AdS BH solutions~\cite{8000,8001,8002,8003,8004,8005,Anand:2025btp}.

On the observational front, BH photon spheres (PSs) serve as a geometric diagnostic of the force balance between gravity and electromagnetism that lies at the heart of the WGC. The existence of unstable PSs outside the EH indicates that gravitational attraction does not overwhelmingly dominate gauge repulsion---precisely the regime the WGC requires---while simultaneously confirming the integrity of the horizon and hence the WCCC~\cite{46m,47m,48m,49m,50m}. BH shadows, as directly observed by the Event Horizon Telescope (EHT) for M87$^{*}$ and Sgr~A$^{*}$, place tight constraints on BH parameters in the strong-field regime, forming a natural bridge between theoretical conjectures and observational data. Gravitational lensing---in both the strong- and weak-deflection limits---offers further observational handles sensitive to the same metric parameters \cite{600,601,602,603,604,605,606,607,608,609,610,611,612,613,614,615}.

The field of observational black hole physics achieved a transformative breakthrough in 2019, when the Event Horizon Telescope (EHT) collaboration released the first horizon-scale image of the supermassive black hole shadow in the galaxy M87* \cite{2029,2030}. This striking observation revealed a bright emission ring surrounding a central dark region—the BH shadow—offering an unprecedented opportunity to test theoretical models of strong-field gravity. The foundational theoretical framework for black hole shadows traces back to Synge \cite{2031}, who demonstrated that a spherically symmetric BH, such as the Schwarzschild solution, produces a perfectly circular shadow. Bardeen later showed that spin distorts this shadow into a D-shaped silhouette in rotating Kerr BHs \cite{2032}. Building on these foundations, Perlick et al. \cite{2033} analytically derived the angular size of BH shadows in spherically symmetric spacetimes surrounded by a non-magnetized plasma. Atamurotov et al. \cite{2034} extended these studies to rotating Kerr BHs embedded in a plasma with a radial power-law density, exploring the combined effects of spin and plasma. Subsequent work investigated whether plasma leaves detectable imprints on shadow shape and size, with significant efforts by several groups \cite{2035,2036}. These ongoing studies bridge theoretical insights with observations and help elucidate the rich interplay between plasma physics and BH shadows within general relativity (GR) and beyond. Also researchers consider the Horndeski BH solution and study its optical properties in a plasma medium \cite{2037}. This class of BH solutions arises from scalar–tensor theories of gravity that extend GR by incorporating non-minimal couplings between a scalar field and curvature invariants. The motivation for selecting the Horndeski BH solution lies in its relevance to probing potential deviations from GR in strong gravitational fields. Furthermore, the rich parameter space of the Horndeski BH solution enables a detailed investigation of how scalar-field interactions influence observable phenomena such as the photon sphere, light deflection, and shadows in a plasma environment.\\

In the present work, we study the interplay of the WGC and WCCC in $F(R)$--Euler--Heisenberg (EH) BHs in both Anti--de Sitter (AdS) and de Sitter (dS) backgrounds. This class of solutions, arising from the coupling of $F(R)$ modified gravity with nonlinear electrodynamics (NED), exhibits a richer horizon structure and thermodynamic behavior compared to standard GR, and provides a well-suited arena for testing swampland conjectures. Our analysis proceeds along five interconnected directions. First, we derive the $F(R)$--EH BH solution and its thermodynamic properties, including the Hawking temperature, Ashtekar--Magnon--Das (AMD) mass, and the first law. Second, we demonstrate the universal entropy--extremality relation within a perturbative deformation framework, supplying thermodynamic evidence for the WGC. Third, we analyze PSs from both geodesic and topological perspectives, computing their topological charges and establishing that the WGC and WCCC hold simultaneously across the parameter space in AdS and dS backgrounds. Fourth, we compute the gravitational deflection angle in the strong- and weak-field regimes, mapping its dependence on the $F(R)$ correction $f_{R_0}$, the EH coupling $\lambda$, and the charge $q$, and identifying the WGC-compatible parameter regions through lensing observables. Fifth, we construct BH shadow images under isotropic, optically thin accretion and examine phase transitions and Joule--Thomson (JT) expansion within the Barrow (BW) entropy framework, connecting thermodynamic stability to the WGC-allowed parameter space. While our analysis is primarily theoretical, the results are shown to be consistent with the order-of-magnitude constraints inferred from current EHT shadow observations. In this sense, observational considerations serve as a qualitative consistency check rather than a direct parameter estimation or exclusion analysis.

The paper is organized as follows. In section~\ref{isec2}, we present the $F(R)$--EH BH solution and its thermodynamic properties. Section~\ref{isec3} establishes the universal relation between entropy and extremality shifts as thermodynamic evidence for the WGC. In section~\ref{isec4}, we study PSs, their topological charges, and the WGC--WCCC compatibility in AdS and dS spacetimes. Section~\ref{isec5} is devoted to gravitational lensing in the strong- and weak-deflection regimes. BH shadows with isotropic accretion are treated in section~\ref{isec6}. Phase transitions and JT expansion within the BW entropy framework are examined in section~\ref{isec7}. We conclude in section~\ref{isec8} with a summary of the main results, observational constraints from EHT data, and an outlook on future directions.

\section{$F(R)$--Euler--Heisenberg BH Solution and Thermodynamics}\label{isec2}

In this section, we present the static, spherically symmetric BH solution that emerges from the interplay of $F(R)$ modified gravity and EH NED, and derive its basic thermodynamic quantities. Earlier work on BHs with nonlinear electromagnetic fields and $F(R)$ corrections---including their PSs, shadow formation, and evaporation processes---can be found in~\cite{a',b'}. The theoretical construction rests on the requirement that the energy--momentum tensor remains traceless~\cite{c',d'}.

\subsection{Blackening function and horizon structure}\label{isec2a}

The spacetime geometry is taken to be static and spherically symmetric, with line element
\begin{equation}\label{e1}
ds^{2} = -h(r)\, dt^{2} + \frac{dr^{2}}{h(r)} + r^{2}\big(d\theta^{2} + \sin^{2}\theta\, d\varphi^{2}\big),
\end{equation}
where \(h(r)\) denotes the metric function, which we refer to as the blackening function. Assuming constant scalar curvature $R = R_{0}$, the $F(R)$ trace equation~\cite{e'} yields
\begin{equation}\label{eq1}
R_{0}(1 + f_{R_{0}}) - 2 \big(R_{0} + \tilde{f}(R_{0})\big) = 0, 
\end{equation}
with $f_{R_{0}} = d\tilde{f}(R)/dR\big|_{R=R_{0}}$, so that the modified gravity sector is characterized by the two parameters $R_{0}$ and $f_{R_{0}}$. The resulting field equations read
\begin{equation}\label{eq3}
R_{\mu\nu}(1 + f_{R_{0}}) - \tfrac{1}{4}\, g_{\mu\nu}\, R_{0}(1 + f_{R_{0}}) = 8\pi T_{\mu\nu}.
\end{equation}
For the EH nonlinear electromagnetic field, the field-strength tensor and its invariants are
\begin{equation}\label{eq4}
P_{\mu\nu} = \frac{q}{r^{2}} \big(\delta^{0}_{\mu}\delta^{1}_{\nu} - \delta^{0}_{\nu}\delta^{1}_{\mu}\big), \qquad
P = \frac{q^{2}}{2r^{4}},\quad O = 0,
\end{equation}
where $q$ is the electric charge parameter. Substituting into the field equations and solving, one obtains the blackening function~\cite{f'}
\begin{equation}\label{eq7}
h(r) = 1 - \frac{m_{0}}{r} - \frac{R_{0}\, r^{2}}{12} + \frac{1}{1+f_{R_{0}}}\!\left(\frac{q^{2}}{r^{2}} - \frac{\lambda\, q^{4}}{20\,r^{6}}\right),
\end{equation}
where $m_{0}$ denotes the geometric mass, $\lambda$ the EH coupling constant, and $R_0$ the constant scalar curvature. Setting $f_{R_{0}}=0$ and $R_{0}=4\Lambda$ recovers the standard GR BH with cosmological constant $\Lambda$. Positive $R_{0}$ gives asymptotically dS spacetimes, while negative $R_{0}$ gives AdS. When $\lambda=0$, the EH correction vanishes and the solution reduces to the RN--(A)dS BH.

The horizon structure is governed by the roots of $h(r_{h})=0$. Depending on the values of $q$, $\lambda$, $f_{R_0}$, and $R_0$, the solution admits non-extremal configurations with two distinct horizons, extremal BHs with a degenerate horizon, or naked singularities where no real positive root exists.

The behavior of $h(r)$ is displayed in Fig.~\ref{fig:metric_AdS} for the AdS case ($R_0=-1$) and in Fig.~\ref{fig:metric_dS} for the dS case ($R_0=+1$). In the AdS background (Fig.~\ref{fig:metric_AdS}), the Schwarzschild-AdS limit ($q=0$, $\lambda=0$, $f_{R_0}=0$) produces a single event horizon at $r_h \simeq 0.932$, beyond which $h(r)$ grows monotonically due to the $r^2/12$ AdS contribution. Switching on the charge introduces an inner Cauchy horizon: for instance, $q=0.3$ with $f_{R_0}=-0.1$ yields a non-extremal BH with horizons at $r_h \simeq \{0.113, 0.832\}$, while $q=0.4$ with $f_{R_0}=0$ gives $r_h \simeq \{0.200, 0.752\}$. As $q$ increases further (or $f_{R_0}$ becomes sufficiently positive), the two horizons merge and eventually disappear, producing a naked singularity; the case $q=0.8$ with $f_{R_0}=0.5$ illustrates this regime, where $h(r)$ remains strictly positive for all $r>0$.

In the dS background (Fig.~\ref{fig:metric_dS}), all curves exhibit the characteristic turnover induced by the $-R_0 r^2/12$ term, which drives $h(r)$ negative at large $r$. The Schwarzschild-dS solution ($q=0$) displays two horizons---a BH event horizon at $r_h \simeq 1.116$ and a cosmological horizon at $r_c \simeq 2.769$. The charged case $q=0.3$ with $f_{R_0}=-0.1$ develops a third inner horizon at $r_h \simeq 0.113$, in addition to the BH horizon at $r_h \simeq 0.975$ and the cosmological horizon at $r_c \simeq 2.807$. The three-dimensional diagram of the AdS spatial geometry is shown in Fig.~\ref{fig:3d_embed}, where the event horizon is marked by a red ring and the radial growth of the metric function toward the AdS boundary is apparent in the contour structure.

\begin{figure}[http!]
\centering
\includegraphics[width=0.85\textwidth]{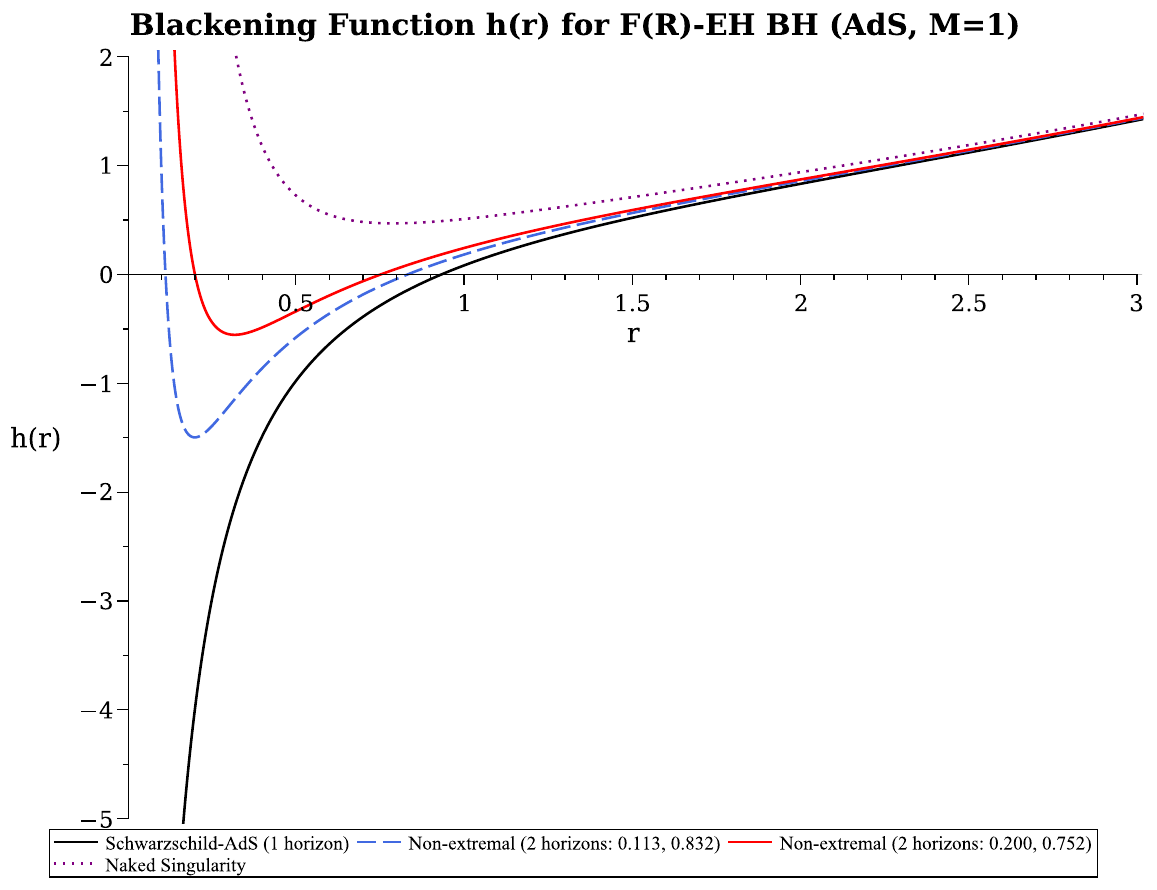}
\caption{Blackening function $h(r)$ for the $F(R)$--EH BH in the AdS background ($R_0 = -1$, $M=1$). The solid black curve corresponds to the Schwarzschild-AdS limit ($q=0$) with a single horizon at $r_h \simeq 0.932$. The blue dashed curve ($q=0.3$, $f_{R_0}=-0.1$) and the red solid curve ($q=0.4$, $f_{R_0}=0$) represent non-extremal configurations with two horizons. The purple dotted curve ($q=0.8$, $f_{R_0}=0.5$) stays entirely above the $h=0$ axis, indicating a naked singularity.}
\label{fig:metric_AdS}
\end{figure}

\begin{figure}[http!]
\centering
\includegraphics[width=0.75\textwidth]{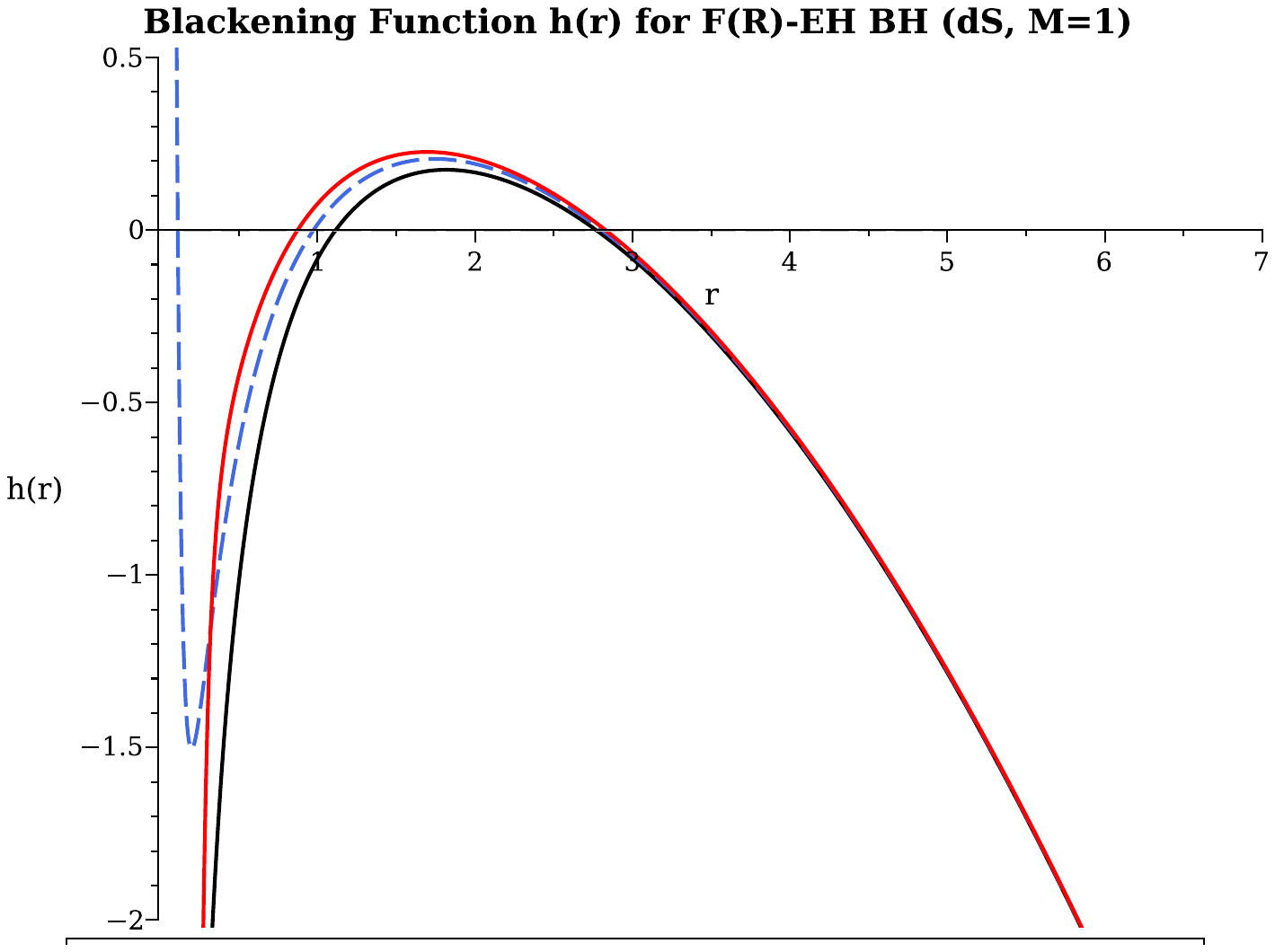}
\caption{Blackening function $h(r)$ for the $F(R)$--EH BH in the dS background ($R_0 = +1$, $M=1$). The solid black curve is the Schwarzschild-dS solution with a BH horizon at $r_h \simeq 1.116$ and a cosmological horizon at $r_c \simeq 2.769$. The blue dashed curve ($q=0.3$, $f_{R_0}=-0.1$) exhibits three horizons: an inner horizon, a BH horizon, and a cosmological horizon. The red solid curve ($q=0.4$, $\lambda=0.5$, $f_{R_0}=0$) shows a two-horizon configuration. All curves turn over and become negative at large $r$, reflecting the dS asymptotic structure.}
\label{fig:metric_dS}
\end{figure}

\begin{figure}[http!]
\centering
\includegraphics[width=0.65\textwidth]{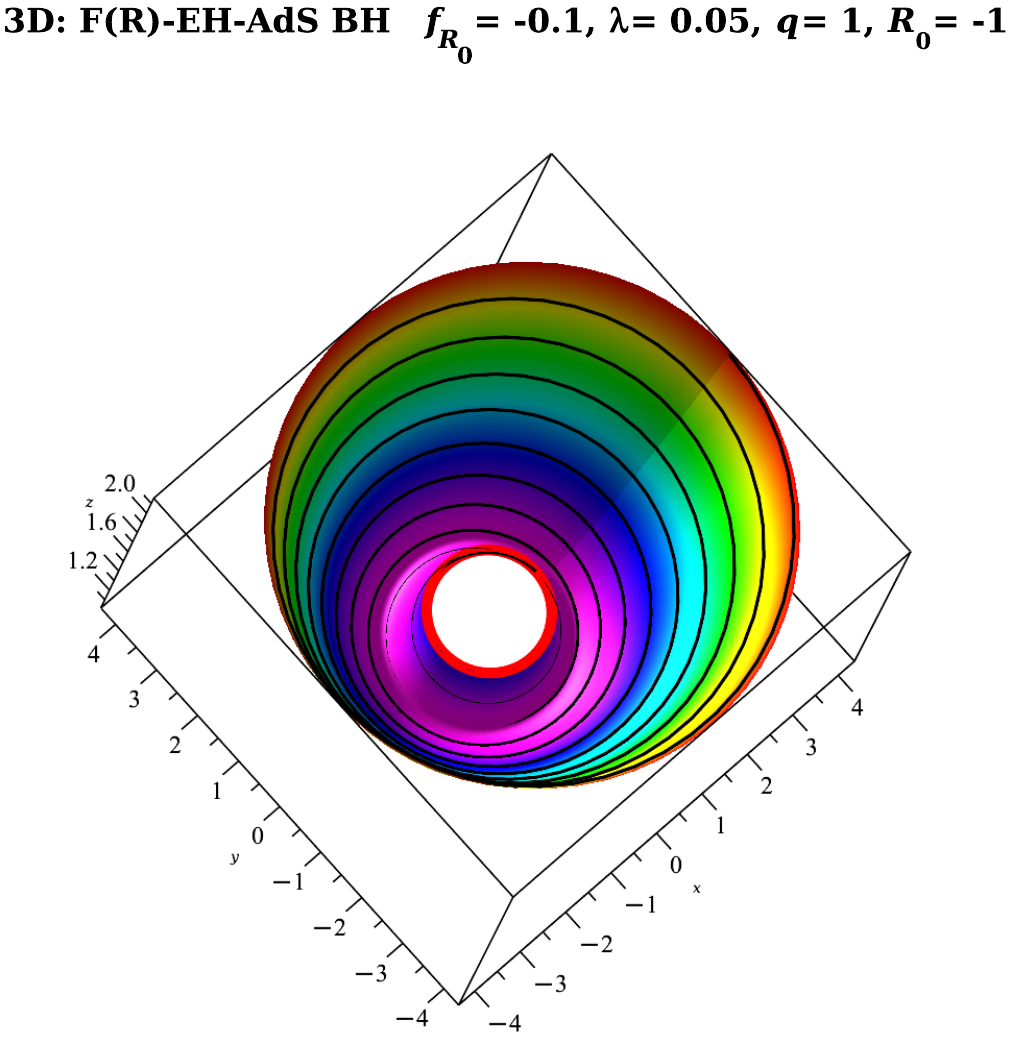}
\caption{Three-dimensional diagram of the $F(R)$--EH-AdS BH spatial geometry for $f_{R_0}=-0.1$, $\lambda=0.05$, $q=1$, and $R_0=-1$. The red ring marks the event horizon. The contour lines trace the monotonic growth of $h(r)$ toward the AdS boundary, and the color gradient encodes the magnitude of the metric function across the radial extent of the spacetime.}
\label{fig:3d_embed}
\end{figure}

\subsection{Hawking temperature}\label{isec2b}

Imposing $h(r_{h})=0$ at the event horizon yields the mass parameter
\begin{equation}\label{eq8}
m_{0} = r_{h} - \frac{R_{0}\, r_{h}^{3}}{12} + \frac{q^{2}}{(1+f_{R_{0}})\, r_{h}}\!\left(1 - \frac{\lambda\, q^{2}}{20\,r_{h}^{4}}\right).
\end{equation}
The Hawking temperature is obtained from the surface gravity $\kappa = \frac{1}{2}|h'(r_{h})|$. A remark on normalization is in order. For spacetimes that are not asymptotically flat, the standard surface gravity formula requires a correction factor tied to the asymptotic value of $g_{tt}$. In the present case, however, this subtlety does not affect the result. For AdS ($R_{0}<0$), the metric function $h(r)$ diverges as $r\to\infty$; the timelike Killing vector $\partial_t$ is normalized at the conformal boundary, and the AMD mass formalism~\cite{g'} already incorporates this prescription. For dS ($R_{0}>0$), a cosmological horizon appears and the Hawking temperature is defined locally at each horizon without reference to spatial infinity. In both cases, the expression
\begin{equation}\label{eq9}
T_{H} = \frac{1}{4\pi}\,h'(r_{h}) 
= \frac{1}{4\pi r_{h}} - \frac{R_{0}\,r_{h}}{16\pi} + \frac{q^{2}}{16\pi\,(1+f_{R_{0}})\, r_{h}^{3}}\!\left(\frac{\lambda\, q^{2}}{r_{h}^{4}} - 4\right)
\end{equation}
remains valid for (A)dS BH thermodynamics, as is standard in the literature~\cite{8001,8002,8003}. We have verified this numerically: the analytic expression~\eqref{eq9} and the numerical derivative $h'(r_h)/(4\pi)$ agree to within $\mathcal{O}(10^{-8})$ across all tested parameter configurations.

The Hawking temperature $T_H(r_h, q)$ is plotted as a three-dimensional surface in Fig.~\ref{fig:TH_AdS} for the AdS case and in Fig.~\ref{fig:TH_dS} for the dS case. In the AdS background (Fig.~\ref{fig:TH_AdS}), the temperature exhibits a non-monotonic dependence on $r_h$: it decreases from a high value at small horizon radii, reaches a local minimum at intermediate $r_h$, and then rises again at large $r_h$ due to the positive contribution $+R_0 r_h/(16\pi)$ from the AdS curvature (recall that $R_0 < 0$ flips the sign). This minimum corresponds to the onset of a Hawking--Page-type transition. The red (clipped) region at small $r_h$ and large $q$ marks the domain where $T_H$ would become negative---signaling configurations inside the inner horizon or beyond the extremal limit, which are not physically realized as BH states. As the charge $q$ grows, the minimum deepens and shifts to larger $r_h$, reflecting the increased repulsive electromagnetic contribution.

In the dS background (Fig.~\ref{fig:TH_dS}), the temperature is a monotonically decreasing function of $r_h$: the dS term $-R_0 r_h/(16\pi)$ (now with $R_0>0$) provides a negative contribution that grows linearly, eventually driving $T_H$ to zero at the cosmological horizon. The surface is dominated by a broad plateau at $T_H \approx 0$ (red region), corresponding to large BH horizons approaching the Nariai limit where the BH and cosmological horizons merge and the temperature vanishes.

\begin{figure}[http!]
\centering
\includegraphics[width=0.65\textwidth]{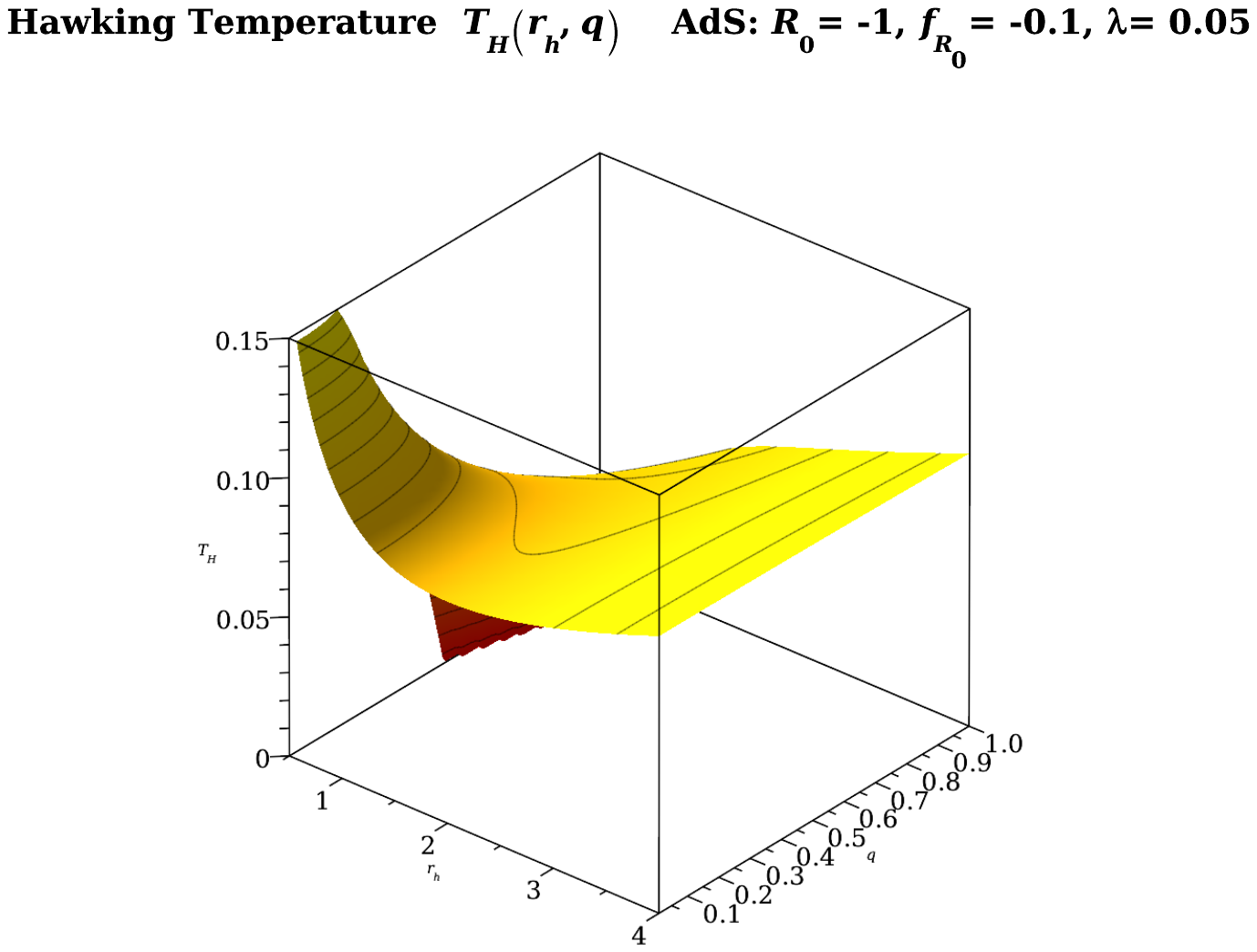}
\caption{Three-dimensional surface of the Hawking temperature $T_H(r_h, q)$ for the $F(R)$--EH-AdS BH with $R_0=-1$, $f_{R_0}=-0.1$, and $\lambda=0.05$. The temperature is non-monotonic in $r_h$: it decreases to a local minimum before rising due to the AdS curvature contribution. The red region corresponds to $T_H \leq 0$ (clipped), which marks configurations beyond the extremal limit.}
\label{fig:TH_AdS}
\end{figure}

\begin{figure}[http!]
\centering
\includegraphics[width=0.65\textwidth]{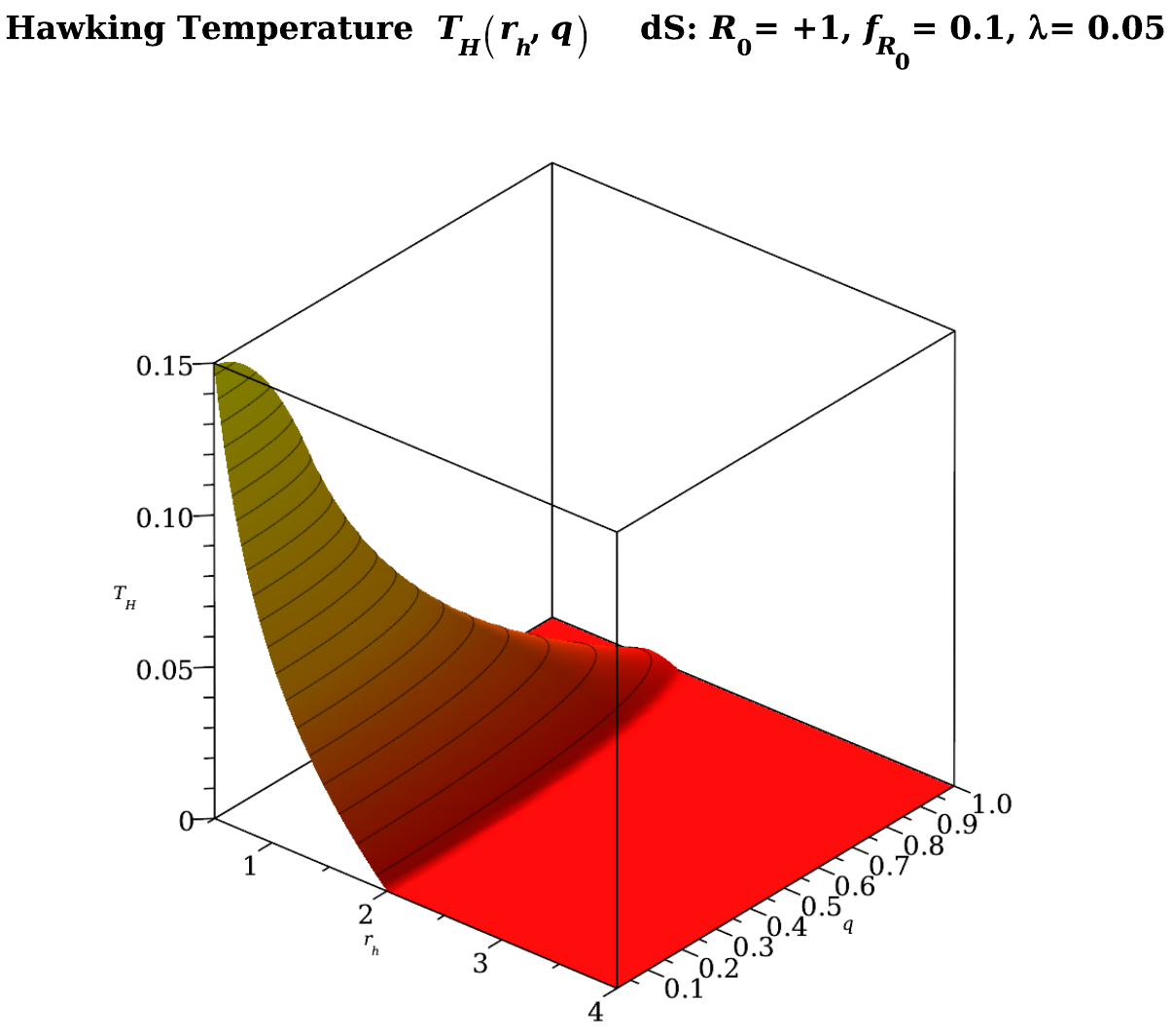}
\caption{Three-dimensional surface of the Hawking temperature $T_H(r_h, q)$ for the $F(R)$--EH-dS BH with $R_0=+1$, $f_{R_0}=0.1$, and $\lambda=0.05$. The temperature decreases monotonically with $r_h$ and vanishes at the Nariai limit. The extensive red plateau at $T_H=0$ reflects the large domain in which the BH horizon approaches the cosmological horizon.}
\label{fig:TH_dS}
\end{figure}

\subsection{AMD mass and first law}\label{isec2c}

Applying the AMD formalism~\cite{g'}, the total mass is
\begin{equation}\label{eq10}
M = \frac{m_{0}(1+f_{R_{0}})}{2} = -\frac{(1+f_{R_{0}})}{24}\, r_{h}(R_{0}\, r_{h}^{2} - 12) - \frac{\lambda\, q^{4} - 20\, q^{2}\, r_{h}^{4}}{40\, r_{h}^{5}}.
\end{equation}
These thermodynamic quantities satisfy the first law
\begin{equation}\label{eq11}
dM = T_{H}\, dS + \Phi\, dQ,
\end{equation}
with $S = \pi r_h^2$ the Bekenstein--Hawking entropy and $\Phi$ the electric potential at the horizon.

The AMD mass $M(r_h, q)$ is shown in Fig.~\ref{fig:mass_AdS} for the AdS case and in Fig.~\ref{fig:mass_dS} for the dS case. In the AdS background (Fig.~\ref{fig:mass_AdS}), the mass is a monotonically increasing function of $r_h$ at fixed $q$, consistent with the thermodynamic stability of large AdS BHs. At small $r_h$ and large $q$, a sharp spike appears due to the $q^4/(40\,r_h^5)$ EH correction, which dominates in the near-extremal regime. The mass grows approximately as $r_h^3$ at large horizon radii, reflecting the volume scaling associated with the AdS curvature through the term $(1+f_{R_0})\,R_0\,r_h^3/24$.

In the dS background (Fig.~\ref{fig:mass_dS}), the mass surface exhibits qualitatively different behavior. It rises to a maximum at intermediate $r_h$ and subsequently decreases, eventually becoming negative at large horizon radii. This turnover is a direct consequence of the dS contribution $-(1+f_{R_0})\,R_0\,r_h^3/24$, which now has the opposite sign and drives the mass downward. The maximum of $M(r_h, q)$ corresponds to the Nariai configuration, beyond which the BH and cosmological horizons merge and no static BH solution exists. The non-monotonicity of the mass function has direct implications for the thermodynamic stability analysis carried out in section~\ref{isec7}.

\begin{figure}[http!]
\centering
\includegraphics[width=0.65\textwidth]{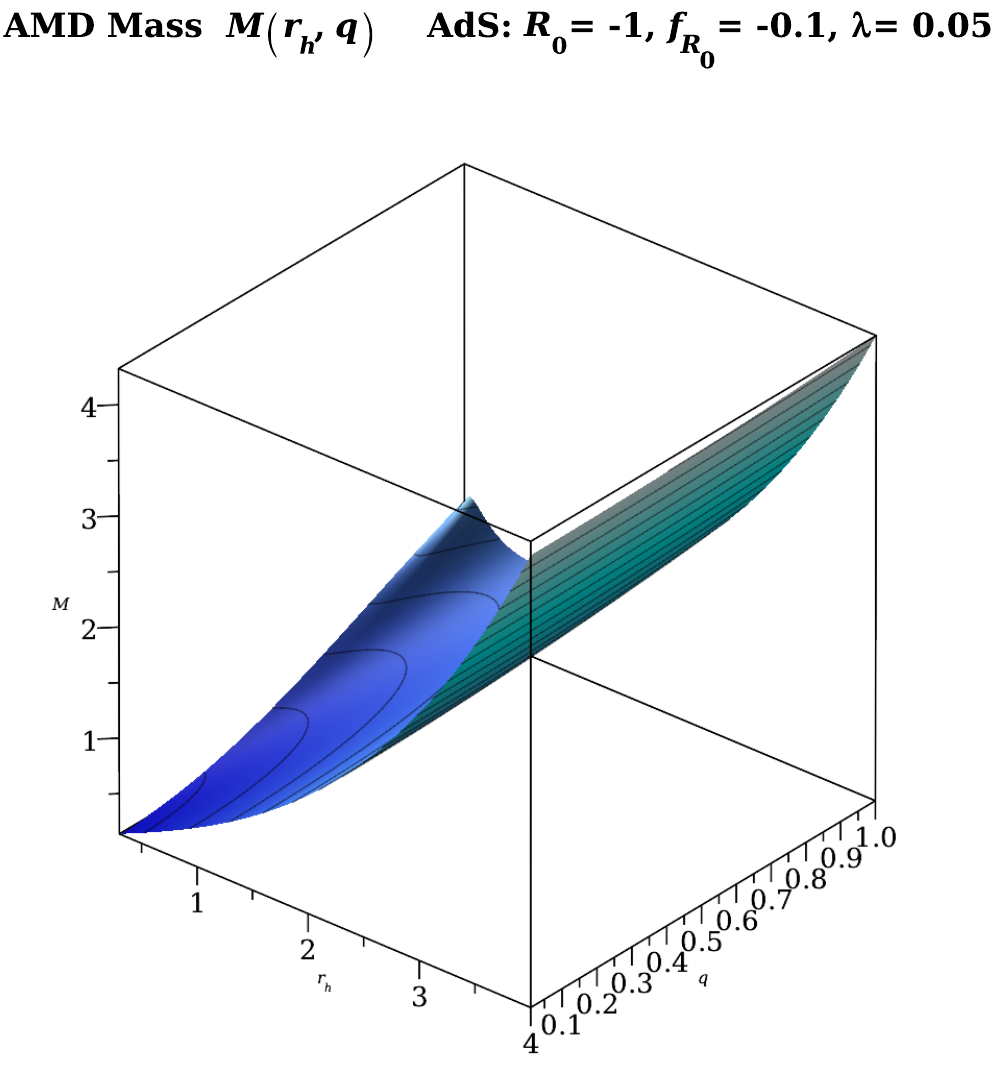}
\caption{Three-dimensional surface of the AMD mass $M(r_h, q)$ for the $F(R)$--EH-AdS BH with $R_0=-1$, $f_{R_0}=-0.1$, and $\lambda=0.05$. The mass grows monotonically with $r_h$, approximately as $r_h^3$ at large radii. The spike at small $r_h$ and large $q$ originates from the EH correction term $\lambda q^4/(40\,r_h^5)$.}
\label{fig:mass_AdS}
\end{figure}

\begin{figure}[http!]
\centering
\includegraphics[width=0.65\textwidth]{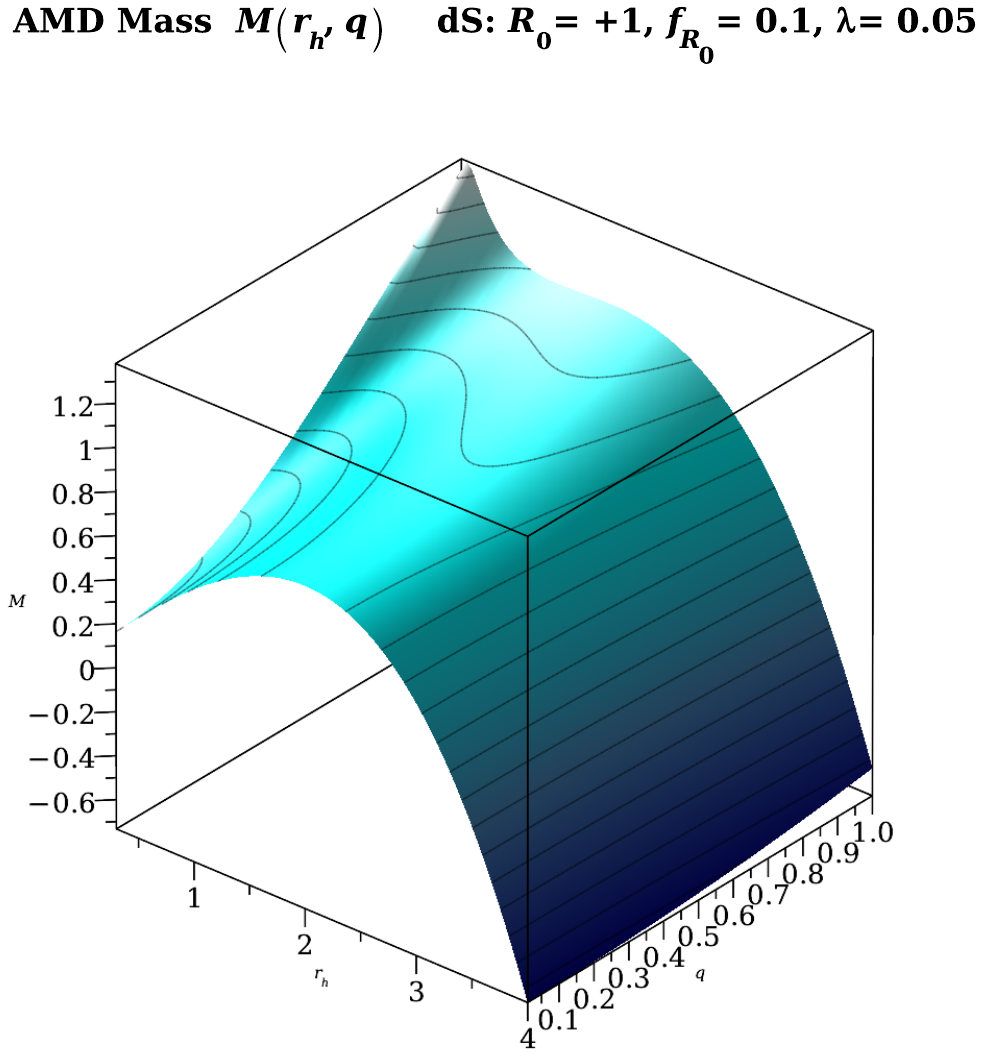}
\caption{Three-dimensional surface of the AMD mass $M(r_h, q)$ for the $F(R)$--EH-dS BH with $R_0=+1$, $f_{R_0}=0.1$, and $\lambda=0.05$. Unlike the AdS case, the mass is non-monotonic: it reaches a maximum at intermediate $r_h$ (corresponding to the Nariai limit) and decreases at larger radii, eventually becoming negative. This behavior is driven by the sign reversal of the dS curvature contribution.}
\label{fig:mass_dS}
\end{figure}

\section{WGC: Universal Relation and Extremality}\label{isec3}

In this section, we establish thermodynamic evidence for the WGC by computing the universal relation between entropy corrections and extremal mass shifts within a perturbative deformation framework. The method follows the approach developed by Goon and Penco~\cite{Goon:2019faz} and subsequently extended to various (A)dS BH backgrounds~\cite{8001,8002,8003,8004,8005,Anand:2025btp}.

\subsection{Perturbative deformation framework}\label{isec3a}

We consider a one-parameter family of deformations of the blackening function~\eqref{eq7}, parameterized by a continuous variable $\eta$ that shifts the EH coupling $\lambda \to \lambda + \eta$. The deformed blackening function reads
\begin{equation}\label{eq12}
h(r;\eta) = 1 - \frac{m_0(\eta)}{r} - \frac{R_0\, r^2}{12} + \frac{1}{1+f_{R_0}}\!\left(\frac{q^2}{r^2} - \frac{(\lambda + \eta)\, q^4}{20\, r^6}\right),
\end{equation}
where the geometric mass $m_0(\eta)$ adjusts to maintain the horizon condition $h(r_h;\eta)=0$. At the extremal point, the horizon condition and the vanishing of the surface gravity are simultaneously imposed:
\begin{equation}\label{eq13}
h(r_{\text{ext}};\eta) = 0, \qquad h'(r_{\text{ext}};\eta) = 0.
\end{equation}
These two conditions determine the extremal horizon radius $r_{\text{ext}}(\eta)$ and the extremal mass parameter $m_{0,\text{ext}}(\eta)$ as functions of the deformation parameter.

\subsection{Entropy--extremality relation}\label{isec3b}

Differentiating the horizon condition $h = 0$ with respect to $\eta$ at fixed charge $q$ yields
\begin{equation}\label{eq14a}
h'(r_{\text{ext}})\,\frac{\partial r_{\text{ext}}}{\partial\eta} + \frac{\partial h}{\partial m_0}\,\frac{\partial m_{0,\text{ext}}}{\partial\eta} + \frac{\partial h}{\partial\eta} = 0.
\end{equation}
At extremality $h'(r_{\text{ext}}) = 0$, so the first term vanishes. With $\partial h/\partial m_0 = -1/r$ and $\partial h/\partial\eta = -q^4/\big(20(1+f_{R_0})\,r^6\big)$, one obtains
\begin{equation}\label{eq14b}
\frac{\partial m_{0,\text{ext}}}{\partial\eta} = -\frac{q^4}{20\,(1+f_{R_0})\, r_{\text{ext}}^5}.
\end{equation}
Since the AMD mass is $M_{\text{ext}} = m_{0,\text{ext}}(1+f_{R_0})/2$, the factor $(1+f_{R_0})$ cancels exactly, yielding the relation
\begin{equation}\label{eq14}
{\frac{\partial M_{\text{ext}}}{\partial\eta} = -\frac{q^4}{40\,r_{\text{ext}}^5} = -\frac{q^4\,\pi^{5/2}}{40\,S^{5/2}},}
\end{equation}
where $S = \pi r_{\text{ext}}^2$ is the Bekenstein--Hawking entropy of the extremal BH.

Several features of this result deserve comment. First, the right-hand side of Eq.~\eqref{eq14} is strictly negative for all $q > 0$, independently of the sign of $R_0$ or the value of $f_{R_0}$. This means the extremal mass decreases under the EH deformation $\eta > 0$, regardless of whether the background is AdS or dS. In the WGC language, the deformation lowers the extremality bound: a BH that was previously extremal at mass $M$ now becomes super-extremal, opening a decay channel. This is precisely the condition demanded by the WGC---no stable extremal remnant should persist in a consistent quantum gravity theory.

Second, the cancellation of $f_{R_0}$ from the final expression~\eqref{eq14} is noteworthy. The $F(R)$ correction modifies both the geometric mass $m_0$ through the horizon condition and the AMD mass $M$ through the prefactor $(1+f_{R_0})/2$; these two effects compensate exactly when the deformation acts on the EH sector. The extremal mass shift is therefore determined entirely by the charge $q$ and the extremal entropy $S$, with no memory of the $F(R)$ modification.

Third, expressing the relation in terms of the entropy makes the connection to information-theoretic arguments transparent. As $S$ increases, the magnitude of $\partial M_{\text{ext}}/\partial\eta$ decreases as $S^{-5/2}$: large BHs are less sensitive to the EH deformation, while small (near-Planckian) BHs experience the strongest extremality shift. This is consistent with the expectation that UV corrections become negligible in the IR limit.

\subsection{Numerical verification}\label{isec3c}

To confirm the relation~\eqref{eq14}, we solve the system~\eqref{eq13} simultaneously for $(r_{\text{ext}}, m_{0,\text{ext}})$ at each value of $q$, and evaluate $\partial M_{\text{ext}}/\partial\eta$ by a centered finite difference with step $\delta\eta = 10^{-7}$, seeding the perturbed solver within $1\%$ of the unperturbed extremal radius to ensure branch continuity.

The AdS results ($R_0=-1$, $f_{R_0}=-0.1$, $\lambda=0.05$) are collected in Table~\ref{tab:WGC_AdS}. The numerical and analytic values of $\partial M_{\text{ext}}/\partial\eta$ agree to within $\mathcal{O}(10^{-3})$ or better across all tested charge values $q \in [0.28, 0.60]$. The dS results ($R_0=+1$, $f_{R_0}=0.1$, $\lambda=0.05$) are presented in Table~\ref{tab:WGC_dS}, where the BH extremal branch is tracked for $q \in [0.32, 0.60]$; agreement of the same order is obtained. In both cases, the derivative $\partial M_{\text{ext}}/\partial\eta$ is strictly negative and grows in magnitude with $q$, confirming that the EH deformation destabilizes the extremal state as required by the WGC. The fact that the analytic formula $-q^4/(40\,r_{\text{ext}}^5)$ reproduces the numerical data to $\mathcal{O}(10^{-3})$ for $f_{R_0} = -0.1$ (AdS) and $f_{R_0} = 0.1$ (dS) simultaneously verifies the exact cancellation of $f_{R_0}$ predicted by Eq.~\eqref{eq14}. The graphical comparison is displayed in Fig.~\ref{fig:universal_relation}, where the red analytic curves and blue numerical markers are indistinguishable on the plotted scale.

\begin{table}[http!]
\centering
\caption{Numerical verification of the entropy--extremality relation~\eqref{eq14} for the AdS BH ($R_0=-1$, $f_{R_0}=-0.1$, $\lambda=0.05$). The columns list the charge parameter $q$, the extremal horizon radius $r_{\text{ext}}$, the Bekenstein--Hawking entropy $S_{\text{ext}} = \pi r_{\text{ext}}^2$, the numerical derivative $(\partial M_{\text{ext}}/\partial\eta)_{\text{num}}$ obtained by centered finite differencing with $\delta\eta = 10^{-7}$, the analytic prediction $(\partial M_{\text{ext}}/\partial\eta)_{\text{ana}} = -q^4/(40\,r_{\text{ext}}^5)$, and the relative error.}
\label{tab:WGC_AdS}
\renewcommand{\arraystretch}{1.4}
\setlength{\tabcolsep}{6pt}
\begin{tabular}{c c c c c c}
\hline\hline
$q$ & $r_{\text{ext}}$ & $S_{\text{ext}}$ & $\left(\dfrac{\partial M_{\text{ext}}}{\partial\eta}\right)_{\!\text{num}}$ & $\left(\dfrac{\partial M_{\text{ext}}}{\partial\eta}\right)_{\!\text{ana}}$ & Rel.\ error \\[4pt]
\hline
$0.28$ & $0.2137$ & $0.1435$ & $-0.3445$ & $-0.3445$ & $4.2\times 10^{-6}$ \\
$0.32$ & $0.2164$ & $0.1471$ & $-0.5525$ & $-0.5522$ & $4.8\times 10^{-4}$ \\
$0.36$ & $0.2236$ & $0.1570$ & $-0.7520$ & $-0.7520$ & $2.8\times 10^{-5}$ \\
$0.40$ & $0.2317$ & $0.1687$ & $-0.9585$ & $-0.9585$ & $4.3\times 10^{-5}$ \\
$0.44$ & $0.2401$ & $0.1811$ & $-1.1725$ & $-1.1742$ & $1.5\times 10^{-3}$ \\
$0.48$ & $0.2485$ & $0.1940$ & $-1.3975$ & $-1.3999$ & $1.7\times 10^{-3}$ \\
$0.52$ & $0.2568$ & $0.2072$ & $-1.6350$ & $-1.6357$ & $4.6\times 10^{-4}$ \\
$0.56$ & $0.2650$ & $0.2206$ & $-1.8800$ & $-1.8815$ & $7.8\times 10^{-4}$ \\
$0.60$ & $0.2730$ & $0.2341$ & $-2.1375$ & $-2.1369$ & $2.9\times 10^{-4}$ \\
\hline\hline
\end{tabular}
\end{table}

\begin{table}[t]
\centering
\caption{Same as Table~\ref{tab:WGC_AdS} but for the dS BH extremal branch ($R_0=+1$, $f_{R_0}=0.1$, $\lambda=0.05$).}
\label{tab:WGC_dS}
\renewcommand{\arraystretch}{1.4}
\setlength{\tabcolsep}{6pt}
\begin{tabular}{c c c c c c}
\hline\hline
$q$ & $r_{\text{ext}}$ & $S_{\text{ext}}$ & $\left(\dfrac{\partial M_{\text{ext}}}{\partial\eta}\right)_{\!\text{num}}$ & $\left(\dfrac{\partial M_{\text{ext}}}{\partial\eta}\right)_{\!\text{ana}}$ & Rel.\ error \\[4pt]
\hline
$0.32$ & $0.2373$ & $0.1770$ & $-0.3485$ & $-0.3482$ & $9.6\times 10^{-4}$ \\
$0.36$ & $0.2338$ & $0.1717$ & $-0.6015$ & $-0.6016$ & $1.3\times 10^{-4}$ \\
$0.40$ & $0.2390$ & $0.1795$ & $-0.8200$ & $-0.8205$ & $6.7\times 10^{-4}$ \\
$0.44$ & $0.2459$ & $0.1900$ & $-1.0415$ & $-1.0416$ & $5.5\times 10^{-5}$ \\
$0.48$ & $0.2534$ & $0.2018$ & $-1.2700$ & $-1.2698$ & $1.7\times 10^{-4}$ \\
$0.52$ & $0.2611$ & $0.2141$ & $-1.5075$ & $-1.5068$ & $4.9\times 10^{-4}$ \\
$0.56$ & $0.2688$ & $0.2269$ & $-1.7525$ & $-1.7530$ & $2.6\times 10^{-4}$ \\
$0.60$ & $0.2764$ & $0.2400$ & $-2.0125$ & $-2.0085$ & $2.0\times 10^{-3}$ \\
\hline\hline
\end{tabular}
\end{table}

\begin{figure}[t]
\centering
\subfigure[AdS: $R_0=-1$, $f_{R_0}=-0.1$, $\lambda=0.05$]{\includegraphics[width=0.48\textwidth]{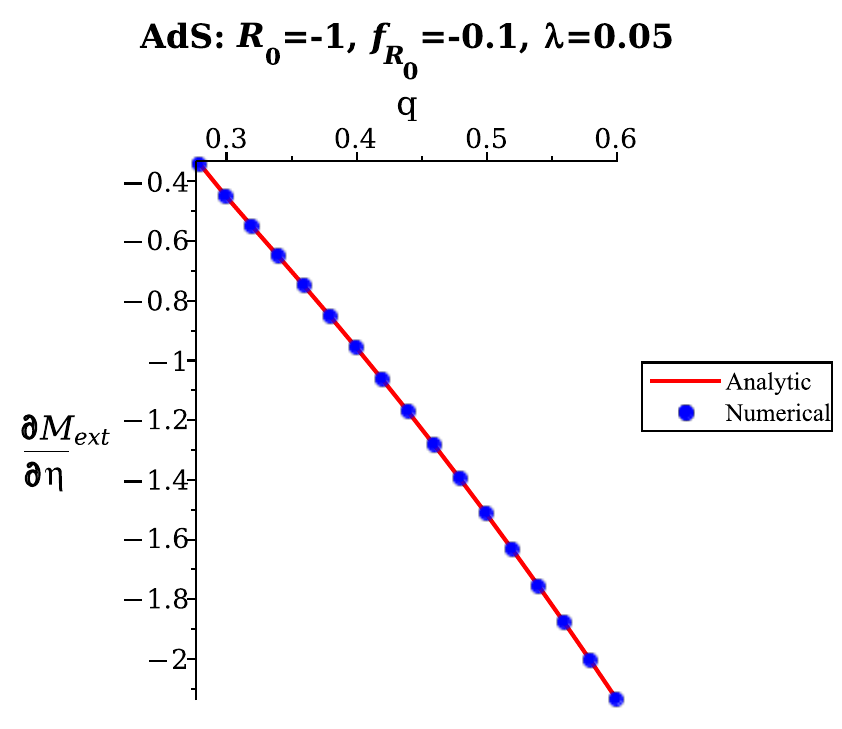}\label{fig:WGC_AdS}}
\hfill
\subfigure[dS: $R_0=+1$, $f_{R_0}=0.1$, $\lambda=0.05$]{\includegraphics[width=0.48\textwidth]{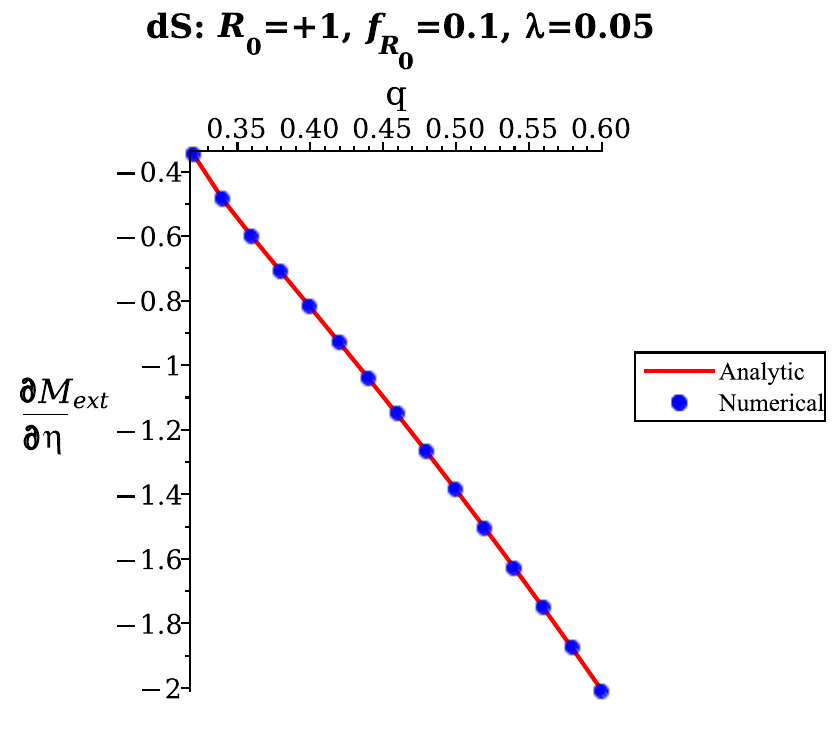}\label{fig:WGC_dS}}
\caption{Graphical comparison of the analytic prediction~\eqref{eq14} (red curves) and the numerical finite-difference evaluation (blue markers) of $\partial M_{\text{ext}}/\partial\eta$ as a function of the charge $q$. Both panels use $\delta\eta=10^{-7}$. The analytic and numerical results are indistinguishable on the plotted scale, confirming the entropy--extremality relation and its independence from the $F(R)$ parameter $f_{R_0}$.}
\label{fig:universal_relation}
\end{figure}

\section{Photon Spheres and WGC--WCCC Compatibility}\label{isec4}

In this section, we analyze the PS structure of the $F(R)$--EH BH from both geodesic and topological perspectives, and use the results to establish the simultaneous compatibility of the WGC and WCCC across the parameter space. The topological classification of PSs follows the framework developed in~\cite{Cunha1,Cunha2,Cunha3} and applied to charged BHs in~\cite{aa',bb',cc',dd',ee',ff',gg',hh',ii',jj'}.

\subsection{Photon sphere condition}\label{isec4a}

For the static, spherically symmetric line element~\eqref{e1}, null geodesics in the equatorial plane ($\theta = \pi/2$) satisfy the effective potential condition for unstable circular orbits:
\begin{equation}\label{eq15}
2\,h(r_{\text{ps}}) - r_{\text{ps}}\, h'(r_{\text{ps}}) = 0,
\end{equation}
where $r_{\text{ps}}$ denotes the PS radius. Substituting the blackening function~\eqref{eq7} and differentiating, the condition becomes
\begin{equation}\label{eq16}
2 - \frac{3\,m_0}{r} - \frac{R_0\, r^2}{4} + \frac{1}{1+f_{R_0}}\!\left(\frac{-2\,q^2}{r^2} + \frac{2\,\lambda\, q^4}{5\,r^6}\right) = 0.
\end{equation}
For given values of the parameters $(m_0, R_0, q, \lambda, f_{R_0})$, this equation admits one or more positive real roots. In the standard RN limit ($\lambda = 0$, $f_{R_0} = 0$), the PS radius reduces to the well-known expression $r_{\text{ps}} = \frac{1}{2}(3m_0 + \sqrt{9m_0^2 - 8q^2})$. The EH correction introduces a $q^4/r^6$ term that shifts the PS inward at small $r$ and can create additional critical points when $\lambda$ is large.

The PS is unstable (i.e., corresponds to a local maximum of the effective potential) when
\begin{equation}\label{eq17}
\frac{d^2}{dr^2}\!\left(\frac{h(r)}{r^2}\right)\bigg|_{r=r_{\text{ps}}} < 0.
\end{equation}
This condition is satisfied for all parameter configurations admitting a PS outside the EH, as confirmed numerically.
To assign a topological charge to each PS, we introduce the potential function~\cite{Cunha1,Cunha2,Cunha3}
\begin{equation}\label{eq18}
H(r,\theta) = \sqrt{\frac{-g_{tt}}{g_{\varphi\varphi}}} = \frac{\sqrt{h(r)}}{r\,\sin\theta},
\end{equation}
whose critical points in the radial direction, $\partial H/\partial r = 0$, coincide with the PS condition~\eqref{eq15}. A two-dimensional vector field is then constructed on the $(r,\theta)$ plane:
\begin{equation}\label{eq19}
\varphi^r_H = \sqrt{h(r)}\,\frac{\partial H}{\partial r}, \qquad \varphi^\theta_H = \frac{1}{r}\,\frac{\partial H}{\partial\theta}.
\end{equation}
The angular component evaluates to
\begin{equation}\label{eq20}
\varphi^\theta_H = -\frac{\sqrt{h(r)}}{r^2}\,\frac{\cos\theta}{\sin^2\theta},
\end{equation}
which diverges at the poles $\theta = 0$ and $\theta = \pi$, ensuring that the vector field is outward-directed at the boundary of the domain $\theta \in (0,\pi)$.

The unit vector field $n^a_H = \varphi^a_H / \lVert \boldsymbol{\varphi}_H\rVert$ ($a = 1,2$ corresponding to $r$ and $\theta$) defines a map from the $(r,\theta)$ plane to the unit circle. The topological current
\begin{equation}\label{eq21}
j^\mu = \frac{1}{2\pi}\,\varepsilon^{\mu\nu\rho}\,\varepsilon_{ab}\,\partial_\nu n^a_H\,\partial_\rho n^b_H
\end{equation}
is conserved ($\partial_\mu j^\mu = 0$) and is nonzero only at the zeros of $\boldsymbol{\varphi}_H$, which are precisely the PSs. The winding number at each zero $z_i$ is
\begin{equation}\label{eq22}
\omega_i = \frac{1}{2\pi}\oint_{\mathcal{C}_i} d\Theta_H,
\end{equation}
where $\Theta_H = \arctan(\varphi^\theta_H / \varphi^r_H)$ and $\mathcal{C}_i$ is a small contour enclosing $z_i$. The total topological charge is $W = \sum_i \omega_i$.

For the $F(R)$--EH BH, each PS is located at $\theta = \pi/2$ (equatorial plane), and the radial component $\varphi^r_H$ changes sign from positive to negative as $r$ increases through $r_{\text{ps}}$, while $\varphi^\theta_H$ points toward the equator from both sides. This circulation pattern yields a winding number $\omega = -1$ at every PS, signaling that all PSs of this spacetime are unstable. The total topological charge is therefore $W = -1$ per PS, consistent with the standard BH classification~\cite{Cunha1,Cunha2,Cunha3}.

\subsection{WGC and PSs in AdS space}\label{isec4c}

The normalized vector field in the $(r,\theta)$ plane is first analyzed for the AdS background ($R_0 = -1$) with $f_{R_0} = -0.1$ and $q_{\text{ext}} = 1$, for three values of the EH coupling: $\lambda = 0.01$, $0.05$, and $0.1$. The corresponding extremal radii are $r_{\text{ext}} = 0.95057359$, $0.93872657$, and $0.93872657$, while the extremal masses take the values $M_{\text{ext}} = 0.98564403$, $0.984338$, and $0.982654$, respectively. For all three configurations, the unit vector field exhibits isolated zero points corresponding to unstable PSs with topological charge $\omega = -1$, situated outside the EH. This confirms that the BH nature of the solution is preserved throughout the explored parameter range.

With $f_{R_0} = -0.1$ in the AdS case, the charge-to-mass ratio satisfies $q_{\text{ext}}/M_{\text{ext}} > 1$ for all three $\lambda$ values: the ratios are $1.01457$, $1.01592$, and $1.01764$ for $\lambda = 0.01$, $0.05$, and $0.1$, respectively. This increasing trend with $\lambda$ demonstrates that the EH nonlinear correction systematically enhances WGC compatibility. However, for generic $f_{R_0}$ values the situation is more nuanced, and a dedicated scan is presented in Table~\ref{tab:wgc2}.

\begin{table}[t]
\centering
\setlength{\tabcolsep}{5pt}
\renewcommand{\arraystretch}{1.3}
\caption{Consistency conditions of WGC, WCCC, and PS for the dS BH ($R_0 = -1$). Entries marked $\times$ indicate that no valid extremal BH configuration exists for those parameters. The column ``WGC--WCCC'' indicates simultaneous compatibility ($\checkmark$).}
\label{tab:wgc2}
\begin{tabular}{cccccccc}
\hline\hline
$f_{R_0}$ & $\lambda$ & $q_{\text{ext}}$ & $r_{\text{ext}}$ & $M_{\text{ext}}$ & $(q/M)_{\text{ext}}>1$ & WGC--WCCC & PS \\[2pt]
\hline
$-0.1$ & $0.01$ & $1$ & $0.95057$ & $0.98564$ & $1.01457$ & $\checkmark$ & $-1$ \\
$-0.1$ & $0.05$ & $1$ & $0.94549$ & $0.98434$ & $1.01591$ & $\checkmark$ & $-1$ \\
$-0.1$ & $0.1$ & $1$ & $0.93873$ & $0.98265$ & $1.01765$ & $\checkmark$ & $-1$ \\
\hline
$0.1$ & $0.01$ & $1$ & $0.87210$ & $1.08289$ & $0.92346$ & $\times$ & $-1$ \\
$0.1$ & $0.05$ & $1$ & $0.86527$ & $1.08087$ & $0.92518$ & $\times$ & $-1$ \\
$0.1$ & $0.1$ & $1$ & $0.85591$ & $1.07822$ & $0.92745$ & $\times$ & $-1$ \\
\hline
$0$ & $0.01$ & $1$ & $0.90875$ & $1.03545$ & $0.96577$ & $\times$ & $-1$ \\
$0$ & $0.05$ & $1$ & $0.90282$ & $1.03381$ & $0.96730$ & $\times$ & $-1$ \\
$0$ & $0.1$ & $1$ & $0.89483$ & $1.03168$ & $0.96930$ & $\times$ & $-1$ \\
\hline\hline
\end{tabular}
\end{table}

\begin{figure}[t]
\centering
\subfigure[]{\includegraphics[height=5cm,width=5cm]{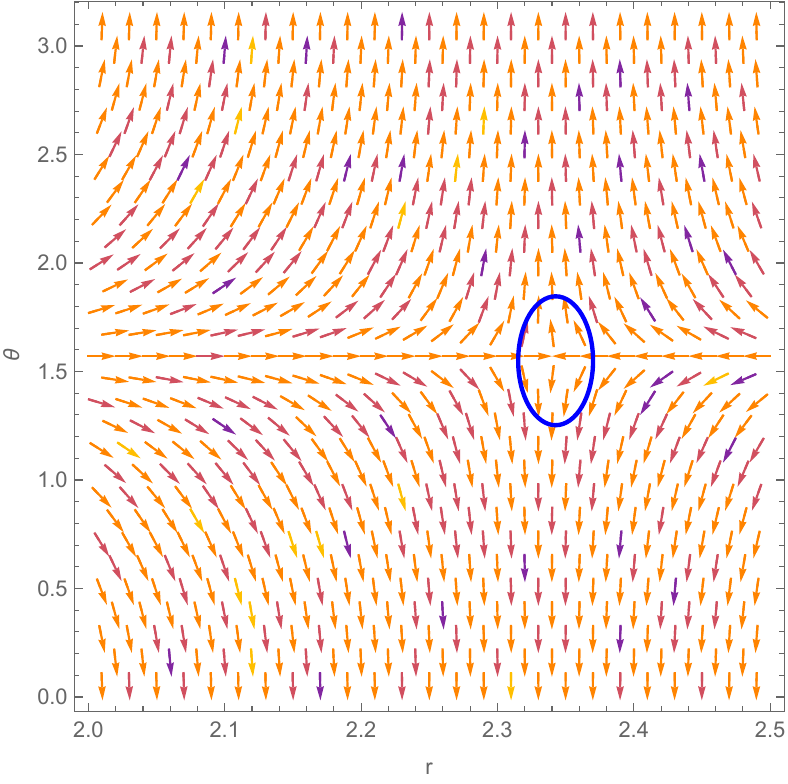}\label{fig:P1}}
\subfigure[]{\includegraphics[height=5cm,width=5cm]{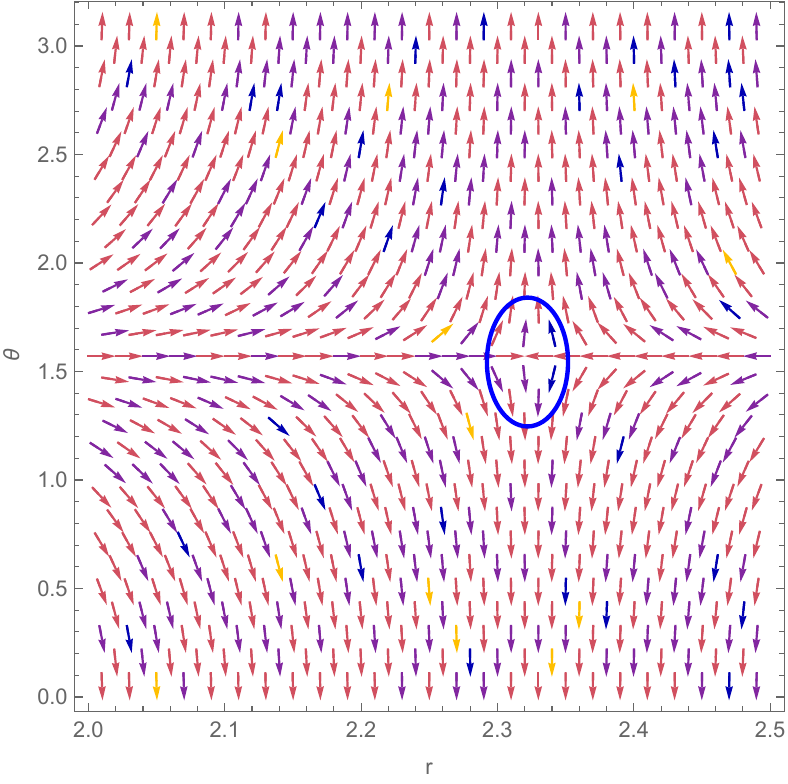}\label{fig:P2}}
\subfigure[]{\includegraphics[height=5cm,width=5cm]{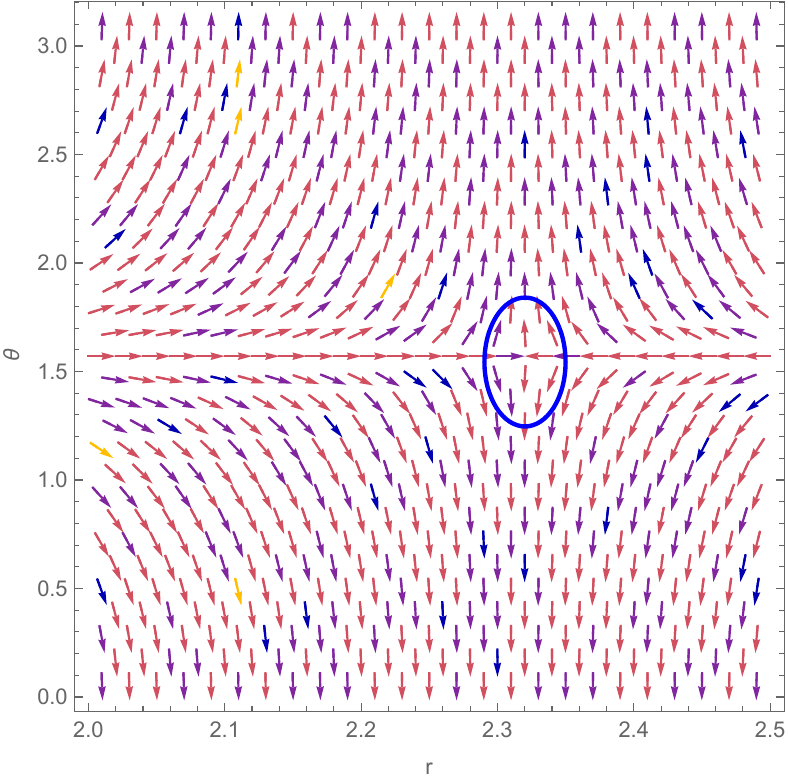}\label{fig:P3}}
\caption{The unit vector field $\vec{n}$ in the $(r,\theta)$ plane for the $F(R)$--EH--AdS BH ($R_0=-1$, $f_{R_0}=-0.1$, $q_{\text{ext}}=1$):
(a)~$\lambda=0.01$, $r_{\text{ext}}=0.95057$, $M_{\text{ext}}=0.98564$;
(b)~$\lambda=0.05$, $r_{\text{ext}}=0.93873$, $M_{\text{ext}}=0.98434$;
(c)~$\lambda=0.1$, $r_{\text{ext}}=0.93873$, $M_{\text{ext}}=0.98265$.
The PS (black dot) carries topological charge $\omega=-1$ in all cases.}
\label{fig:vector_field_AdS}
\end{figure}

\subsection{WGC and PSs in dS space}\label{isec4d}

A similar analysis is carried out for the dS background ($R_0 = +1$). A broader range of parameter choices is examined to assess the robustness of the conclusions. For $f_{R_0} = 0.1$ and $q_{\text{ext}} = 1$, the three EH couplings $\lambda = 0.01$, $0.05$, and $0.1$ yield extremal radii $r_{\text{ext}} = 1.180299$, $1.16506$, and $1.17364$, with associated extremal masses $M_{\text{ext}} = 0.997314$, $0.996871$, and $0.996299$, respectively. In addition, to highlight the role of the $F(R)$ correction, three further cases with $f_{R_0} = 0$ and the same set of $\lambda$ values are analyzed, leading to extremal radii $r_{\text{ext}} = 1.395952$, $1.37154$, and $1.35163$, and extremal masses $M_{\text{ext}} = 0.942763$, $0.942565$, and $0.942297$. In all dS configurations, the unit vector field displays well-defined zero points corresponding to unstable PSs with $\omega = -1$, consistently situated outside the EH.

The results are collected in Table~\ref{tab:wgc1}. Note that the entries with $f_{R_0} = -0.1$ in the dS case do not yield valid extremal BH configurations (marked with $\times$), whereas both $f_{R_0} = 0$ and $f_{R_0} = 0.1$ produce WGC-compatible solutions. For $f_{R_0} = 0$, the charge-to-mass ratios are significantly above unity ($q_{\text{ext}}/M_{\text{ext}} \approx 1.061$), while for $f_{R_0} = 0.1$ they are only marginally above unity ($\approx 1.003$). In all WGC-compatible cases, the PS remains intact with $\omega = -1$, confirming the simultaneous validity of both conjectures.

\begin{table}[t]
\centering
\setlength{\tabcolsep}{5pt}
\renewcommand{\arraystretch}{1.3}
\caption{Consistency conditions of WGC, WCCC, and PS for the dS BH ($R_0 = +1$). Entries marked $\times$ indicate that no valid extremal BH configuration exists for those parameters. The column ``WGC--WCCC'' indicates simultaneous compatibility ($\checkmark$).}
\label{tab:wgc1}
\begin{tabular}{cccccccc}
\hline\hline
$f_{R_0}$ & $\lambda$ & $q_{\text{ext}}$ & $r_{\text{ext}}$ & $M_{\text{ext}}$ & $(q/M)_{\text{ext}}>1$ & WGC--WCCC & PS \\[2pt]
\hline
$-0.1$ & $0.01$ & $1$ & $\times$ & $\times$ & $\times$ & $\times$ & $\times$ \\
$-0.1$ & $0.05$ & $1$ & $\times$ & $\times$ & $\times$ & $\times$ & $\times$ \\
$-0.1$ & $0.1$ & $1$ & $\times$ & $\times$ & $\times$ & $\times$ & $\times$ \\
\hline
$0.1$ & $0.01$ & $1$ & $1.18030$ & $0.99731$ & $1.00269$ & $\checkmark$ & $-1$ \\
$0.1$ & $0.05$ & $1$ & $1.17364$ & $0.99687$ & $1.00314$ & $\checkmark$ & $-1$ \\
$0.1$ & $0.1$ & $1$ & $1.16506$ & $0.99630$ & $1.00371$ & $\checkmark$ & $-1$ \\
\hline
$0$ & $0.01$ & $1$ & $1.39595$ & $0.94276$ & $1.06071$ & $\checkmark$ & $-1$ \\
$0$ & $0.05$ & $1$ & $1.37154$ & $0.94257$ & $1.06094$ & $\checkmark$ & $-1$ \\
$0$ & $0.1$ & $1$ & $1.35163$ & $0.94230$ & $1.06124$ & $\checkmark$ & $-1$ \\
\hline\hline
\end{tabular}
\end{table}

\begin{figure}[t]
\centering
\subfigure[]{\includegraphics[height=5cm,width=5cm]{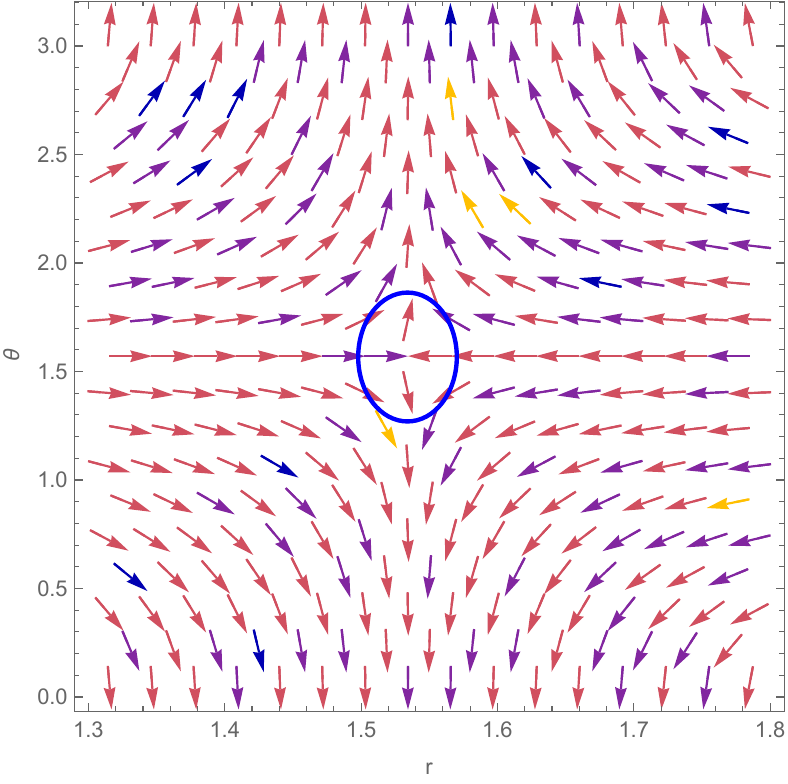}\label{fig:T1}}
\subfigure[]{\includegraphics[height=5cm,width=5cm]{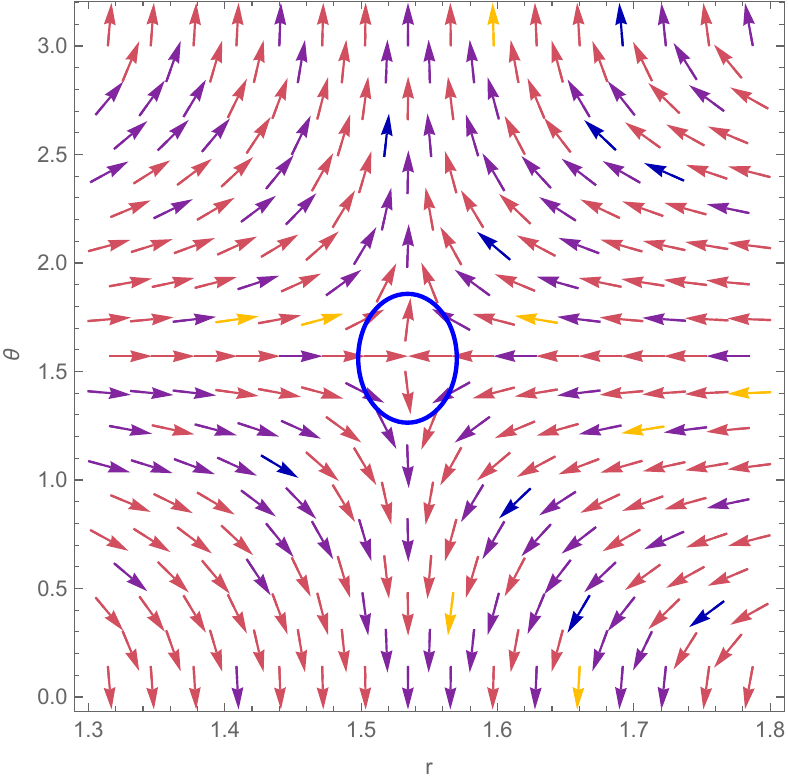}\label{fig:T2}}
\subfigure[]{\includegraphics[height=5cm,width=5cm]{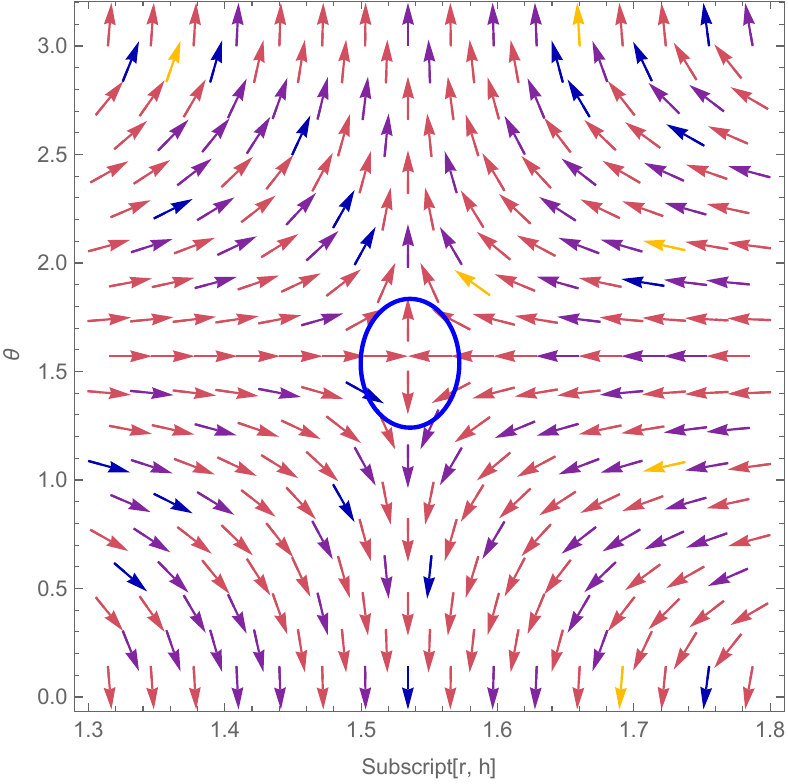}\label{fig:T3}}
\subfigure[]{\includegraphics[height=5cm,width=5cm]{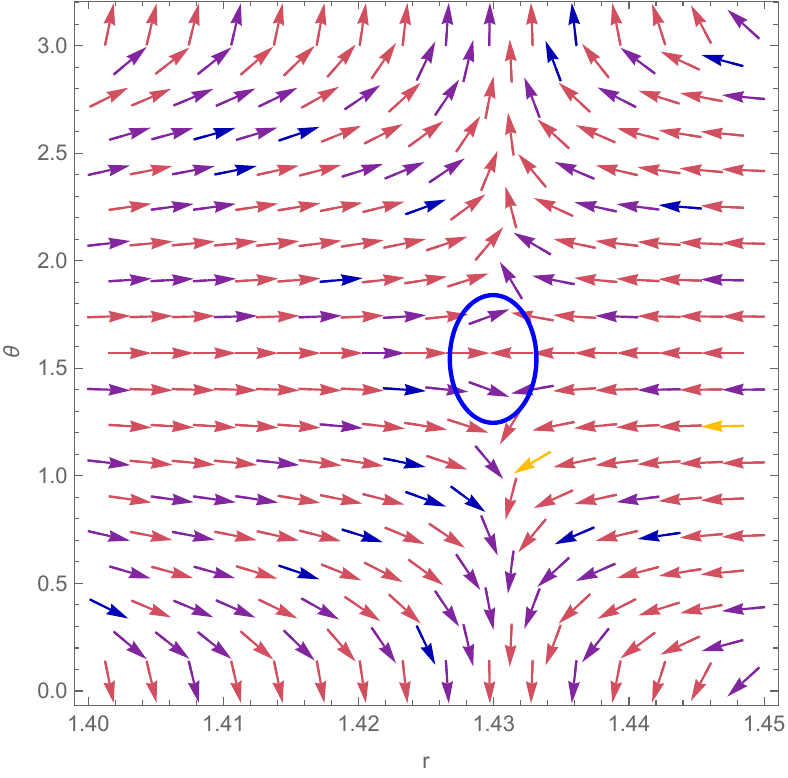}\label{fig:T4}}
\subfigure[]{\includegraphics[height=5cm,width=5cm]{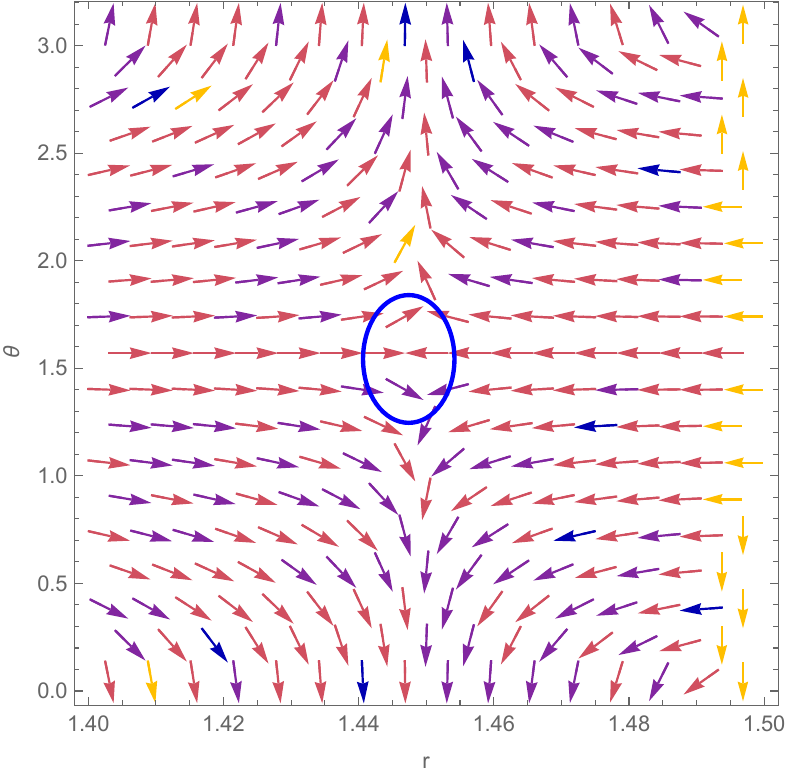}\label{fig:T5}}
\subfigure[]{\includegraphics[height=5cm,width=5cm]{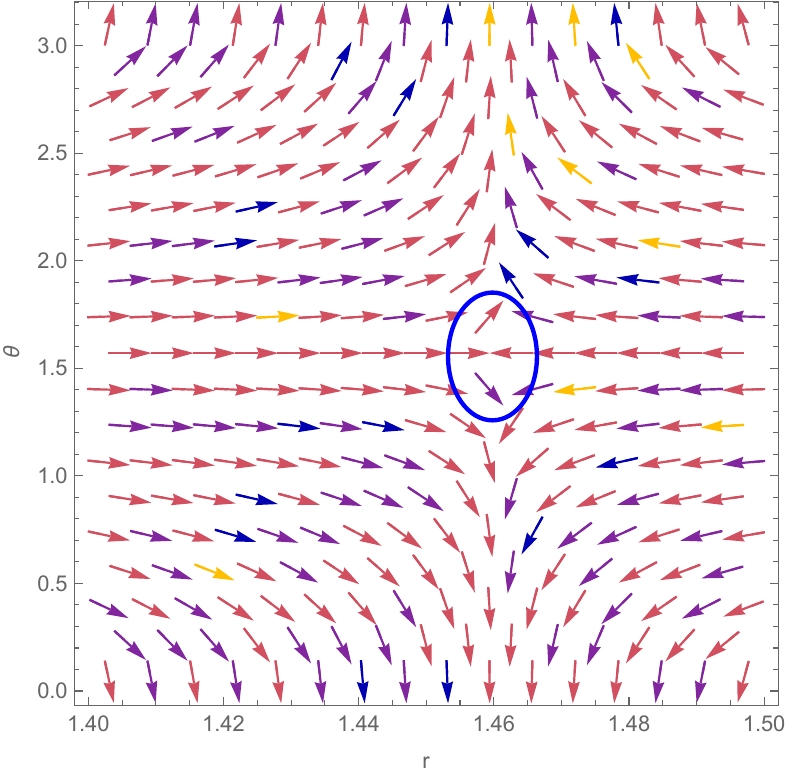}\label{fig:T6}}
\caption{The unit vector field $\vec{n}$ in the $(r,\theta)$ plane for the $F(R)$--EH--dS BH ($R_0=+1$, $q_{\text{ext}}=1$):
(a)~$f_{R_0}=0.1$, $\lambda=0.01$, $r_{\text{ext}}=1.18030$, $M_{\text{ext}}=0.99731$;
(b)~$f_{R_0}=0.1$, $\lambda=0.05$, $r_{\text{ext}}=1.16506$, $M_{\text{ext}}=0.99687$;
(c)~$f_{R_0}=0.1$, $\lambda=0.1$, $r_{\text{ext}}=1.17364$, $M_{\text{ext}}=0.99630$;
(d)~$f_{R_0}=0$, $\lambda=0.01$, $r_{\text{ext}}=1.39595$, $M_{\text{ext}}=0.94276$;
(e)~$f_{R_0}=0$, $\lambda=0.05$, $r_{\text{ext}}=1.37154$, $M_{\text{ext}}=0.94257$;
(f)~$f_{R_0}=0$, $\lambda=0.1$, $r_{\text{ext}}=1.35163$, $M_{\text{ext}}=0.94230$.
In all panels, the PS (black dot) has topological charge $\omega=-1$.}
\label{fig:vector_field_dS}
\end{figure}

\subsection{Discussion}\label{isec4e}

Taken together, the AdS and dS analyses demonstrate that unstable PSs persist across a wide region of the parameter space of $F(R)$--EH BHs. The existence of PSs outside the EH excludes the formation of naked singularities, thereby confirming WCCC compliance. Simultaneously, the charge-to-mass ratio at extremality exceeds unity in all configurations where a well-defined extremal BH exists, satisfying the WGC.

The physical mechanism underlying this compatibility can be traced to the $\lambda q^4/r^6$ term in the blackening function~\eqref{eq7}. This term becomes dominant at small $r$, effectively stiffening the near-horizon geometry and allowing the BH to sustain a larger charge before becoming extremal. As $\lambda$ increases, $M_{\text{ext}}$ decreases monotonically (see Tables~\ref{tab:wgc2} and~\ref{tab:wgc1}), pushing the ratio $q_{\text{ext}}/M_{\text{ext}}$ further above unity and widening the WGC-compatible window. The PS, which probes the geometry at $r_{\text{ps}} > r_h$, is only mildly affected by the EH correction, ensuring that its topological character ($\omega = -1$) remains unchanged throughout.

The $F(R)$ modification, through $f_{R_0}$, provides an additional handle. In the dS background, $f_{R_0} = 0$ yields the strongest WGC signal ($q/M \approx 1.061$), $f_{R_0} = 0.1$ gives marginal compatibility ($q/M \approx 1.003$), while $f_{R_0} = -0.1$ fails to produce valid extremal configurations altogether. This hierarchy reflects the role of the $1/(1+f_{R_0})$ prefactor in $h(r)$: smaller values of $(1+f_{R_0})$ amplify the electromagnetic contribution, but excessively small values can destabilize the horizon structure. The optimal WGC compatibility is thus achieved at intermediate values of $f_{R_0}$, providing a concrete realization of how modified gravity parameters influence swampland conjectures. Once the existence and topological instability of the photon sphere are established, further variations of parameters do not alter its qualitative nature. The role of the Euler–Heisenberg coupling is therefore not to change the topology of photon spheres, but to ensure their persistence outside the event horizon in regimes close to extremality.

\section{Gravitational Lensing}\label{isec5}

In this section, we compute the gravitational deflection angle of the $F(R)$--EH BH in both the strong- and weak-field regimes. This analysis connects the near-horizon geometry studied in sections~\ref{isec3} and~\ref{isec4} to observational signatures accessible through lensing measurements, and maps the dependence of the deflection angle on the parameters $f_{R_0}$, $\lambda$, and $q$. Strong-field lensing follows the Bozza formalism~\cite{Bozza1,Bozza2,Bozza3}, while the weak-field analysis employs the Gauss--Bonnet theorem (GBT) method~\cite{2016}.

\subsection{Strong deflection limit}\label{isec5a}

For the static, spherically symmetric metric~\eqref{e1}, the deflection angle of a photon with closest approach distance $r_0$ is given by
\begin{equation}\label{eq25}
\hat{\alpha}_D(r_0) = I(r_0) - \pi, \qquad I(r_0) = 2\int_{r_0}^{\infty} \frac{dr}{r^2\,\sqrt{\dfrac{1}{b^2} - \dfrac{h(r)}{r^2}}},
\end{equation}
where $b = r_0/\sqrt{h(r_0)}$ is the impact parameter. As $r_0 \to r_{\text{ps}}$, the integral diverges logarithmically. To extract the divergence, one introduces the variable $z = 1 - r_0/r$ and splits the integrand into a divergent part $f_0(z,r_0)$ and a regular part~\cite{Bozza1,Bozza2,Bozza3}:
\begin{equation}\label{eq26}
I(r_0) = \int_0^1 R(z,r_0)\,f(z,r_0)\,dz,
\end{equation}
where
\begin{equation}\label{eq27}
R(z,r_0) = \frac{2r_0}{(1-z)^2}\,\sqrt{\frac{h(r_0)}{h(r)}}, \qquad f(z,r_0) = \frac{1}{\sqrt{h(r_0)\,r^2 - h(r)\,r_0^2}}\,\frac{r_0}{r^2}.
\end{equation}
Near $z = 0$, the function $f$ diverges as $f(z,r_0) \simeq 1/\sqrt{\alpha_1 z + \alpha_2 z^2}$, where $\alpha_1 \to 0$ as $r_0 \to r_{\text{ps}}$ (by the PS condition~\eqref{eq15}), producing the logarithmic divergence.

In the strong deflection limit ($b \to b_c \equiv r_{\text{ps}}/\sqrt{h(r_{\text{ps}})}$), the deflection angle takes the form~\cite{Bozza1,Bozza2,Bozza3}
\begin{equation}\label{eq29}
{\hat{\alpha}_D(b) = -\bar{a}\,\ln\!\left(\frac{b}{b_c} - 1\right) + \bar{b} + \mathcal{O}(b - b_c),}
\end{equation}
where the strong deflection coefficients are
\begin{equation}\label{eq30}
\bar{a} = \sqrt{\frac{2}{2\,h(r_{\text{ps}}) - r_{\text{ps}}^2\, h''(r_{\text{ps}})}}, \qquad \bar{b} = -\pi + I_R(r_{\text{ps}}) + \bar{a}\,\ln\!\left(\frac{2\,\alpha_2\, r_{\text{ps}}^2}{b_c}\right),
\end{equation}
with $I_R(r_{\text{ps}})$ the regularized integral. For the $F(R)$--EH blackening function~\eqref{eq7}, the second derivative entering $\bar{a}$ reads
\begin{equation}\label{eq31}
h''(r_{\text{ps}}) = -\frac{2\,m_0}{r_{\text{ps}}^3} - \frac{R_0}{6} + \frac{1}{1+f_{R_0}}\!\left(\frac{6\,q^2}{r_{\text{ps}}^4} - \frac{42\,\lambda\, q^4}{20\,r_{\text{ps}}^8}\right).
\end{equation}

\subsubsection*{Lensing observables}

For a lens at distance $D_{OL}$ from the observer, the angular position of the relativistic images, their angular separation, and the relative magnification are~\cite{Bozza1,Bozza2,Bozza3}
\begin{equation}\label{eq33}
\theta_\infty = \frac{b_c}{D_{OL}}, \qquad s = \theta_\infty\,\exp\!\left(\frac{\bar{b} - 2\pi}{\bar{a}}\right), \qquad r_{\text{mag}} = \frac{5\pi}{\bar{a}\,\ln 10}.
\end{equation}
The angular separation $s$ between the outermost relativistic image and the asymptotic image position $\theta_\infty$ is exponentially sensitive to the ratio $\bar{b}/\bar{a}$, making it a powerful probe of the EH and $F(R)$ corrections.

\subsubsection*{AdS background}

The strong deflection coefficients for the AdS BH ($R_0 = -1$, $m_0 = 1$) are collected in Table~\ref{tab:SDL_AdS}. In the uncharged limit ($q=0$), the PS radius $r_{\text{ps}} = 1.5$ and the critical impact parameter $b_c = 2.07846$ are independent of $f_{R_0}$ and $\lambda$, recovering the Schwarzschild--AdS values $\bar{a} = 1$ and $\bar{b} = 1.36765$. For $q = 0.3$, the PS shifts inward and the coefficients deviate from the Schwarzschild values, with the deviation increasing as $f_{R_0}$ becomes more negative (i.e., the electromagnetic contribution is amplified). The EH coupling $\lambda$ produces only a mild effect at $q = 0.3$, since the $\lambda q^4/r^6$ correction is small for moderate charge. At $q = 0.5$, however, the strong-field coefficients change dramatically: $\bar{a}$ increases to $\sim 1.28$--$1.41$ and $\bar{b}$ rises to $\sim 6.4$--$8.6$, reflecting the closer proximity of the PS to the horizon and the steeper gravitational potential. The dependence of $\hat{\alpha}_D(b)$ on the impact parameter near $b_c$ is displayed in Fig.~\ref{fig:SDL_AdS}.

\begin{table}[http!]
\centering
\caption{Strong deflection coefficients for the AdS BH ($R_0=-1$, $m_0=1$). Representative entries from the full parameter scan are shown.}
\label{tab:SDL_AdS}
\renewcommand{\arraystretch}{1.3}
\setlength{\tabcolsep}{4pt}
\begin{tabular}{c c c c c c c}
\hline\hline
$f_{R_0}$ & $\lambda$ & $q$ & $r_{\text{ps}}$ & $b_c$ & $\bar{a}$ & $\bar{b}$ \\[2pt]
\hline
\multicolumn{7}{c}{\textit{Schwarzschild--AdS benchmark}} \\
$0$ & $0$ & $0$ & $1.50000$ & $2.07846$ & $1.00000$ & $1.36765$ \\
\hline
\multicolumn{7}{c}{$f_{R_0} = -0.1$} \\
$-0.1$ & $0$ & $0.3$ & $1.35208$ & $1.97760$ & $1.05964$ & $2.50296$ \\
$-0.1$ & $0.05$ & $0.3$ & $1.35210$ & $1.97761$ & $1.05960$ & $2.50262$ \\
$-0.1$ & $0.10$ & $0.3$ & $1.35212$ & $1.97761$ & $1.05955$ & $2.50228$ \\
\hline
\multicolumn{7}{c}{$f_{R_0} = 0$} \\
$0$ & $0$ & $0.3$ & $1.36847$ & $1.98899$ & $1.05183$ & $2.35660$ \\
$0$ & $0.01$ & $0.5$ & $1.00025$ & $1.73213$ & $1.41298$ & $8.56971$ \\
$0$ & $0.05$ & $0.5$ & $1.00124$ & $1.73245$ & $1.40812$ & $8.51970$ \\
$0$ & $0.10$ & $0.5$ & $1.00246$ & $1.73286$ & $1.40219$ & $8.45861$ \\
\hline
\multicolumn{7}{c}{$f_{R_0} = 0.1$} \\
$0.1$ & $0$ & $0.3$ & $1.38156$ & $1.99806$ & $1.04584$ & $2.24393$ \\
$0.1$ & $0$ & $0.5$ & $1.07856$ & $1.78488$ & $1.28114$ & $6.41765$ \\
$0.1$ & $0.05$ & $0.5$ & $1.07920$ & $1.78510$ & $1.27877$ & $6.39480$ \\
$0.1$ & $0.10$ & $0.5$ & $1.07983$ & $1.78532$ & $1.27644$ & $6.37221$ \\
\hline\hline
\end{tabular}
\end{table}

\begin{figure}[http!]
\centering
\includegraphics[width=0.65\textwidth]{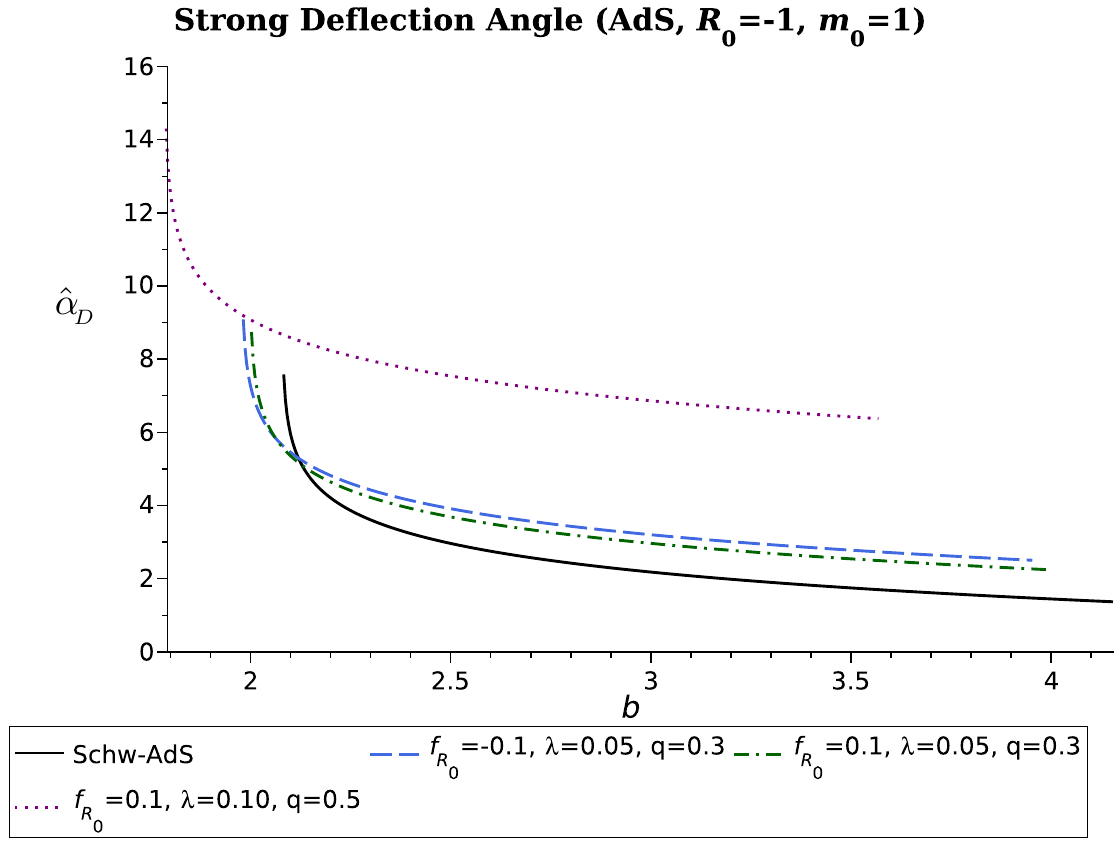}
\caption{Strong deflection angle $\hat{\alpha}_D$ as a function of the impact parameter $b$ near the critical value $b_c$ for the AdS BH ($R_0=-1$, $m_0=1$). The logarithmic divergence as $b \to b_c^+$ is evident. Different curves correspond to varying $(f_{R_0}, \lambda, q)$ as indicated in the legend.}
\label{fig:SDL_AdS}
\end{figure}

\subsubsection*{dS background}

Table~\ref{tab:SDL_dS} presents the corresponding results for the dS BH ($R_0 = +1$). The PS radii $r_{\text{ps}}$ coincide with their AdS counterparts (since the PS condition~\eqref{eq16} is dominated by the near-field terms), but the critical impact parameter $b_c$ is significantly larger---for example, $b_c = 3.92792$ for Schwarzschild--dS versus $2.07846$ for Schwarzschild--AdS---due to the repulsive cosmological contribution that bends light rays outward. The coefficient $\bar{b}$ also increases substantially (e.g., $\bar{b} = 8.97206$ for Schwarzschild--dS versus $1.36765$ for Schwarzschild--AdS), while $\bar{a}$ remains unchanged. Note the entry $f_{R_0} = 0$, $\lambda = 0$, $q = 0.5$ marked ``---'': in this case no PS exists outside the horizon, corresponding to a naked singularity regime. The introduction of even a small $\lambda$ (e.g., $\lambda = 0.01$) restores the PS, demonstrating the stabilizing role of the EH correction. The strong-field deflection angle for the dS case is shown in Fig.~\ref{fig:SDL_dS}.

\begin{table}[http!]
\centering
\caption{Strong deflection coefficients for the dS BH ($R_0=+1$, $m_0=1$). The entry ``---'' indicates no PS outside the horizon.}
\label{tab:SDL_dS}
\renewcommand{\arraystretch}{1.3}
\setlength{\tabcolsep}{4pt}
\begin{tabular}{c c c c c c c}
\hline\hline
$f_{R_0}$ & $\lambda$ & $q$ & $r_{\text{ps}}$ & $b_c$ & $\bar{a}$ & $\bar{b}$ \\[2pt]
\hline
\multicolumn{7}{c}{\textit{Schwarzschild--dS benchmark}} \\
$0$ & $0$ & $0$ & $1.50000$ & $3.92792$ & $1.00000$ & $8.97206$ \\
\hline
\multicolumn{7}{c}{$f_{R_0} = -0.1$} \\
$-0.1$ & $0$ & $0.3$ & $1.35208$ & $3.35145$ & $1.05964$ & $9.77301$ \\
$-0.1$ & $0.05$ & $0.3$ & $1.35210$ & $3.35149$ & $1.05960$ & $9.77301$ \\
$-0.1$ & $0.10$ & $0.3$ & $1.35212$ & $3.35153$ & $1.05955$ & $9.77307$ \\
\hline
\multicolumn{7}{c}{$f_{R_0} = 0$} \\
$0$ & $0$ & $0.5$ & --- & --- & --- & --- \\
$0$ & $0.01$ & $0.5$ & $1.00025$ & $2.44972$ & $1.41298$ & $16.19845$ \\
$0$ & $0.05$ & $0.5$ & $1.00124$ & $2.45063$ & $1.40812$ & $16.13021$ \\
$0$ & $0.10$ & $0.5$ & $1.00246$ & $2.45177$ & $1.40219$ & $16.05846$ \\
\hline
\multicolumn{7}{c}{$f_{R_0} = 0.1$} \\
$0.1$ & $0$ & $0.3$ & $1.38156$ & $3.45404$ & $1.04584$ & $9.57731$ \\
$0.1$ & $0$ & $0.5$ & $1.07856$ & $2.60620$ & $1.28114$ & $13.70754$ \\
$0.1$ & $0.05$ & $0.5$ & $1.07920$ & $2.60688$ & $1.27877$ & $13.67497$ \\
$0.1$ & $0.10$ & $0.5$ & $1.07983$ & $2.60756$ & $1.27644$ & $13.64331$ \\
\hline\hline
\end{tabular}
\end{table}

\begin{figure}[t]
\centering
\includegraphics[width=0.65\textwidth]{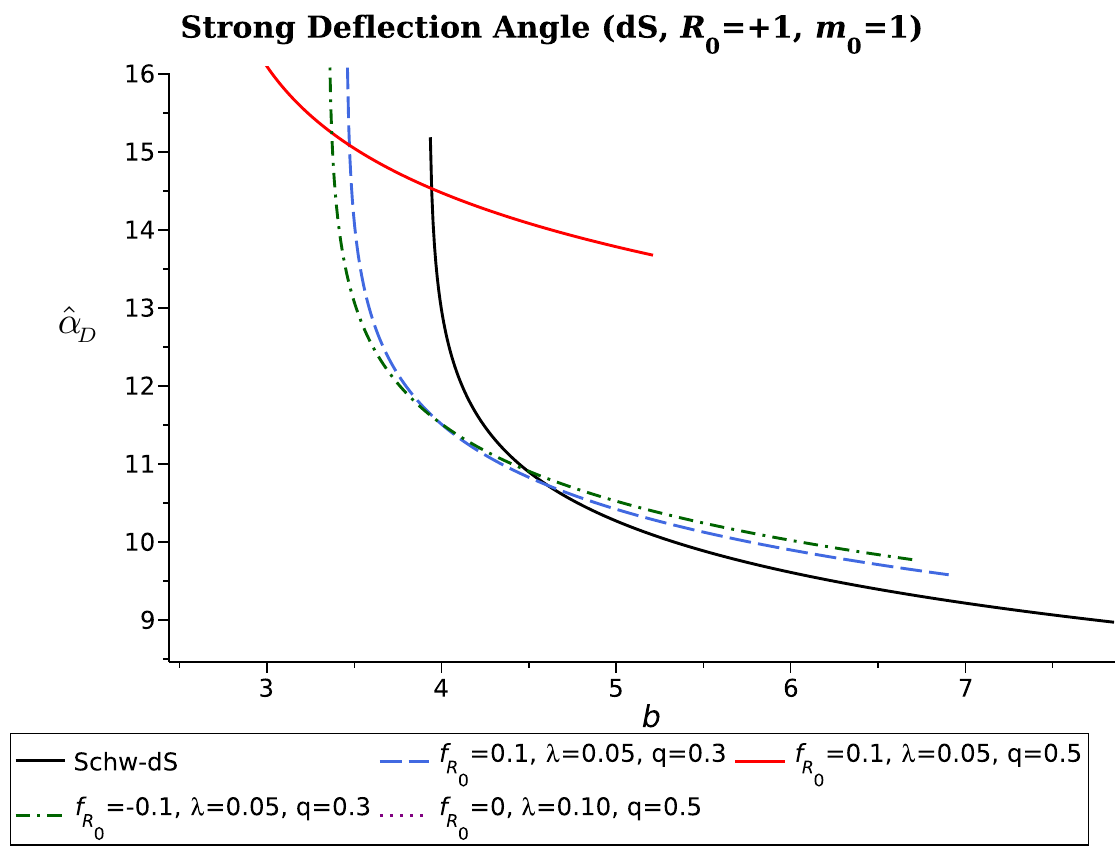}
\caption{Same as Fig.~\ref{fig:SDL_AdS} but for the dS BH ($R_0=+1$). The larger values of $b_c$ and $\bar{b}$ compared to the AdS case are a direct consequence of the positive cosmological constant.}
\label{fig:SDL_dS}
\end{figure}

The key physical trends extracted from Tables~\ref{tab:SDL_AdS} and~\ref{tab:SDL_dS} are: (i)~increasing $q$ shifts the PS inward and increases both $\bar{a}$ and $\bar{b}$, enhancing the strong-field deflection; (ii)~increasing $\lambda$ at fixed $q$ slightly decreases $\bar{a}$ and $\bar{b}$, reflecting the outward shift of the PS caused by the EH stiffening term $\lambda q^4/r^6$; (iii)~more negative $f_{R_0}$ amplifies the electromagnetic contribution, pushing the PS further inward and increasing $\bar{a}$; (iv)~the dS background produces larger $b_c$ and $\bar{b}$ but leaves $\bar{a}$ and $r_{\text{ps}}$ essentially unchanged, confirming that the strong deflection coefficient $\bar{a}$ is a near-field quantity insensitive to the cosmological background.

\subsection{Weak deflection limit via Gauss--Bonnet theorem}\label{isec5b}

In the weak-field regime ($b \gg r_{\text{ps}}$), the deflection angle can be computed using the GBT applied to the optical metric induced by the BH spacetime. For the metric~\eqref{e1}, the optical metric on the equatorial plane ($\theta = \pi/2$) is
\begin{equation}\label{eq34}
d\sigma^2 = \frac{dr^2}{h(r)^2} + \frac{r^2}{h(r)}\,d\varphi^2.
\end{equation}
The Gaussian curvature $\mathcal{K}$ of this two-dimensional Riemannian manifold is
\begin{equation}\label{eq35}
\mathcal{K} = -\frac{1}{\sqrt{g_{\text{opt}}}}\!\left[\frac{\partial}{\partial r}\!\left(\frac{\sqrt{g_{\text{opt}}}}{g_{rr}}\,\Gamma^r_{\varphi\varphi,r}\right)\right],
\end{equation}
where $g_{\text{opt}} = r^2/h(r)^3$ is the determinant of the optical metric. Applying the GBT to a domain bounded by the photon trajectory and a circular arc at infinity yields
\begin{equation}\label{eq36}
\hat{\alpha}_D = -\iint_{\mathcal{D}} \mathcal{K}\,dS.
\end{equation}
For the $F(R)$--EH blackening function~\eqref{eq7}, expanding to leading order and integrating along the straight-line trajectory $r = b/\sin\varphi$, the weak-field deflection angle is
\begin{equation}\label{eq38}
{\hat{\alpha}_D^{\text{weak}} = \frac{4\,m_0}{b} - \frac{3\pi\, q^2}{4\,(1+f_{R_0})\,b^2} + \frac{3\pi\,\lambda\, q^4}{16\,(1+f_{R_0})\,b^6} - \frac{R_0\, b}{6} + \mathcal{O}\!\left(\frac{m_0^2}{b^2}\right).}
\end{equation}
The first term is the standard Schwarzschild deflection. The second is the RN-type correction, modified by the $F(R)$ factor $(1+f_{R_0})$. The third is the EH correction, positive and dominant only at small $b$. The fourth reflects the cosmological background: $R_0 < 0$ (AdS) produces a positive contribution that grows with $b$, while $R_0 > 0$ (dS) gives a negative correction.

\subsubsection*{AdS background}

Table~\ref{tab:WDL} collects the weak deflection angle at representative impact parameters $b = 5$, $10$, $15$, and $20$ for both AdS and dS backgrounds. In the AdS case ($R_0 = -1$), the deflection angle is dominated by the cosmological term $-R_0 b/6 = b/6$ at large $b$, producing monotonically increasing $\hat{\alpha}_D^{\text{weak}}$ as the impact parameter grows. For instance, the Schwarzschild--AdS value ($q=0$) rises from $1.633$ at $b=5$ to $3.533$ at $b=20$. The charge correction is negative and small: at $b=5$ it reduces $\hat{\alpha}_D$ by $\sim 0.026$ for $q=0.5$ with $f_{R_0}=-0.1$, and by $\sim 0.022$ for $f_{R_0}=0.1$. The EH coupling $\lambda$ has negligible effect at these impact parameters, as the $\lambda q^4/b^6$ term is suppressed by the sixth power of $b$. Fig.~\ref{fig:WDL_AdS} displays $\hat{\alpha}_D^{\text{weak}}(b)$ for AdS, with Fig.~\ref{fig:WDL_AdS_zoom} providing a magnified view of the small-$b$ region where the charge and EH corrections are most pronounced.

\subsubsection*{dS background}

In the dS case ($R_0 = +1$), the cosmological term $-b/6$ renders the deflection angle negative for all but the smallest impact parameters. The Schwarzschild--dS value ranges from $-0.033$ at $b=5$ to $-3.133$ at $b=20$. The charge correction deepens the deflection: $q=0.5$ with $f_{R_0}=-0.1$ gives $-0.060$ at $b=5$ and $-3.135$ at $b=20$. As in the AdS case, $\lambda$ has negligible impact at large $b$. The negative deflection at large impact parameters is a hallmark of dS spacetimes, arising from the competition between gravitational attraction and cosmological repulsion. Figs.~\ref{fig:WDL_dS} and~\ref{fig:WDL_dS_zoom} display the full and magnified views, respectively.

\begin{table}[http!]
\centering
\caption{Weak deflection angle (in radians) at selected impact parameters for AdS ($R_0=-1$) and dS ($R_0=+1$) backgrounds with $m_0=1$.}
\label{tab:WDL}
\renewcommand{\arraystretch}{1.3}
\setlength{\tabcolsep}{4pt}
\begin{tabular}{c c c c c c c c}
\hline\hline
$R_0$ & $f_{R_0}$ & $\lambda$ & $q$ & $b=5$ & $b=10$ & $b=15$ & $b=20$ \\[2pt]
\hline
\multicolumn{8}{c}{\textit{AdS background}} \\
$-1$ & $-0.1$ & $0.05$ & $0$ & $1.6333$ & $2.0667$ & $2.7667$ & $3.5333$ \\
$-1$ & $-0.1$ & $0.05$ & $0.3$ & $1.6239$ & $2.0643$ & $2.7656$ & $3.5327$ \\
$-1$ & $-0.1$ & $0.05$ & $0.5$ & $1.6072$ & $2.0601$ & $2.7638$ & $3.5317$ \\
$-1$ & $0$ & $0.05$ & $0.5$ & $1.6098$ & $2.0608$ & $2.7641$ & $3.5319$ \\
$-1$ & $0.1$ & $0.05$ & $0.5$ & $1.6119$ & $2.0613$ & $2.7643$ & $3.5320$ \\
$-1$ & $-0.1$ & $0.10$ & $0.5$ & $1.6072$ & $2.0601$ & $2.7638$ & $3.5317$ \\
\hline
\multicolumn{8}{c}{\textit{dS background}} \\
$+1$ & $0.1$ & $0.05$ & $0$ & $-0.0333$ & $-1.2667$ & $-2.2333$ & $-3.1333$ \\
$+1$ & $0.1$ & $0.05$ & $0.3$ & $-0.0410$ & $-1.2686$ & $-2.2342$ & $-3.1338$ \\
$+1$ & $0.1$ & $0.05$ & $0.5$ & $-0.0548$ & $-1.2720$ & $-2.2357$ & $-3.1347$ \\
$+1$ & $0$ & $0.05$ & $0.5$ & $-0.0569$ & $-1.2726$ & $-2.2360$ & $-3.1348$ \\
$+1$ & $-0.1$ & $0.05$ & $0.5$ & $-0.0595$ & $-1.2732$ & $-2.2362$ & $-3.1350$ \\
$+1$ & $0$ & $0.10$ & $0.5$ & $-0.0569$ & $-1.2726$ & $-2.2360$ & $-3.1348$ \\
\hline\hline
\end{tabular}
\end{table}

\begin{figure}[http!]
\centering
\subfigure[Full range]{\includegraphics[width=0.48\textwidth]{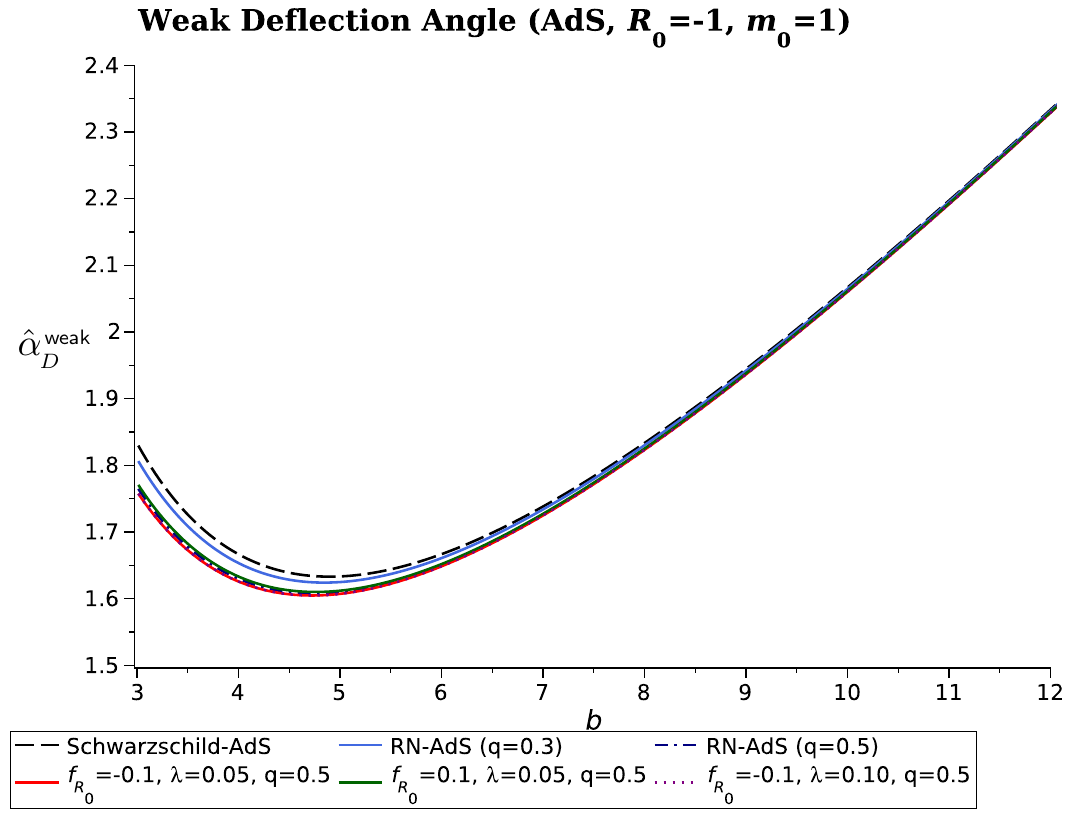}\label{fig:WDL_AdS}}
\hfill
\subfigure[Magnified near small $b$]{\includegraphics[width=0.48\textwidth]{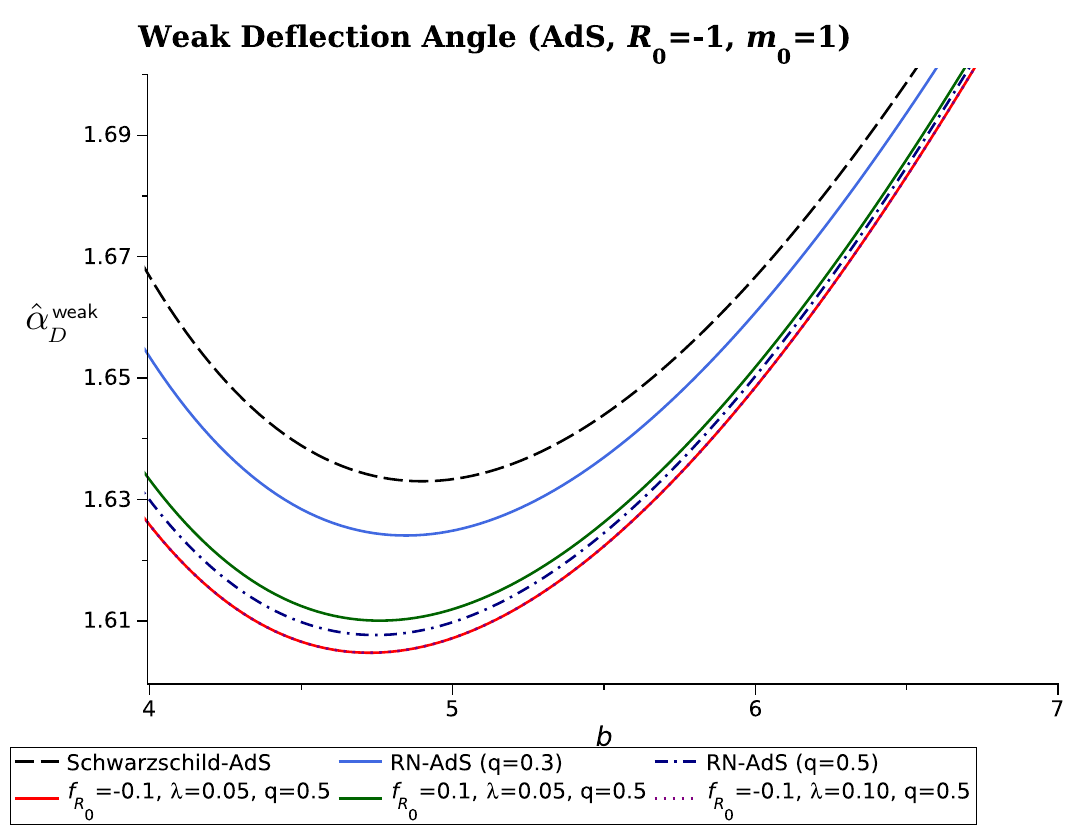}\label{fig:WDL_AdS_zoom}}
\caption{Weak deflection angle $\hat{\alpha}_D^{\text{weak}}$ versus impact parameter $b$ for the AdS BH ($R_0=-1$, $m_0=1$). Panel~(a) shows the full range; panel~(b) provides a magnified view of the small-$b$ region where the charge and EH corrections are most visible.}
\label{fig:WDL_AdS_panels}
\end{figure}

\begin{figure}[http!]
\centering
\subfigure[Full range]{\includegraphics[width=0.48\textwidth]{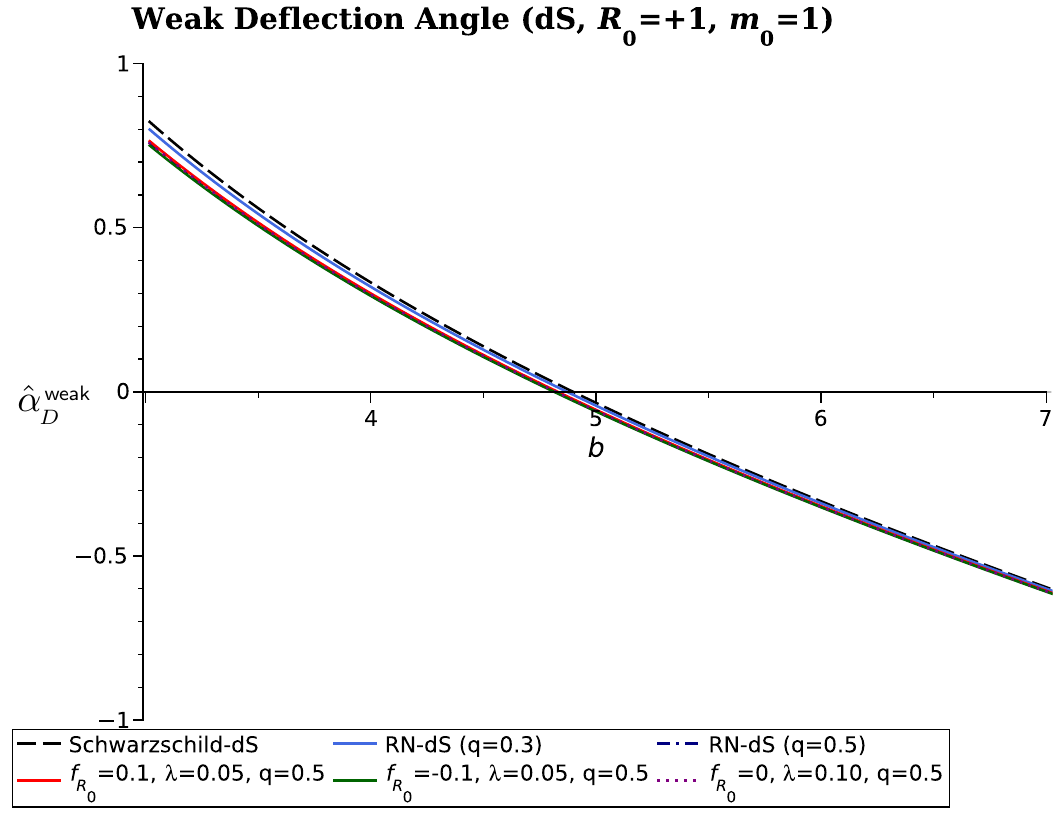}\label{fig:WDL_dS}}
\hfill
\subfigure[Magnified near small $b$]{\includegraphics[width=0.48\textwidth]{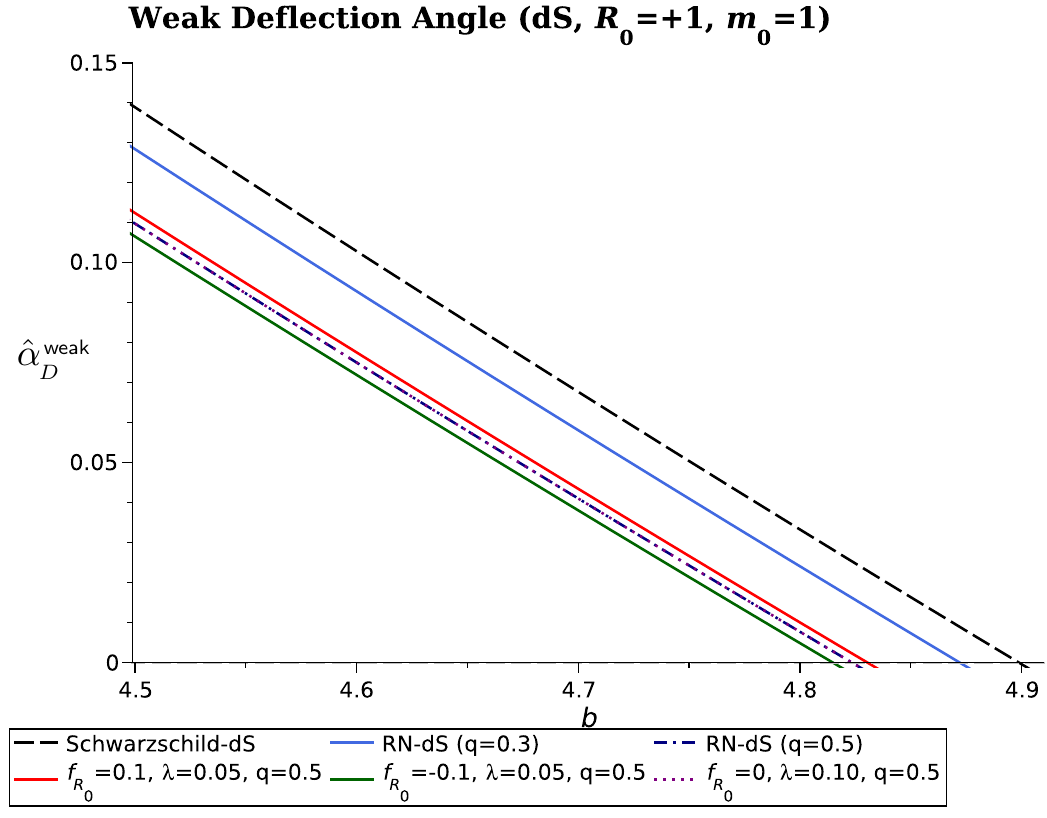}\label{fig:WDL_dS_zoom}}
\caption{Same as Fig.~\ref{fig:WDL_AdS_panels} but for the dS BH ($R_0=+1$). The deflection angle becomes increasingly negative at large $b$ due to the positive cosmological constant.}
\label{fig:WDL_dS_panels}
\end{figure}

\subsubsection*{Parameter dependence}

The role of the individual parameters is isolated in Figs.~\ref{fig:WDL_fR0_panels} and~\ref{fig:WDL_lam_panels}. Fig.~\ref{fig:WDL_fR0} displays the effect of $f_{R_0}$ at fixed $q = 0.5$ and $\lambda = 0.05$ in the AdS background: more negative $f_{R_0}$ enhances the electromagnetic deflection through the factor $1/(1+f_{R_0})$, producing a larger (more positive) shift. This is the same mechanism that governs WGC compatibility in section~\ref{isec4}: the parameter space region where the WGC is satisfied ($q_{\text{ext}}/M_{\text{ext}} > 1$) corresponds to enhanced electromagnetic deflection, providing a lensing signature of WGC compatibility.

Fig.~\ref{fig:WDL_lam} isolates the EH coupling $\lambda$ at fixed $q = 0.5$ and $f_{R_0} = -0.1$. While the $\lambda$-dependence is barely perceptible at the scale of the full plot, the zoomed view (Fig.~\ref{fig:WDL_lam_zoom}) reveals the expected trend: larger $\lambda$ increases $\hat{\alpha}_D^{\text{weak}}$ at small $b$ through the positive $\lambda q^4/b^6$ term, though the effect remains subdominant compared to the charge and cosmological contributions.

\begin{figure}[http!]
\centering
\subfigure[Full range]{\includegraphics[width=0.48\textwidth]{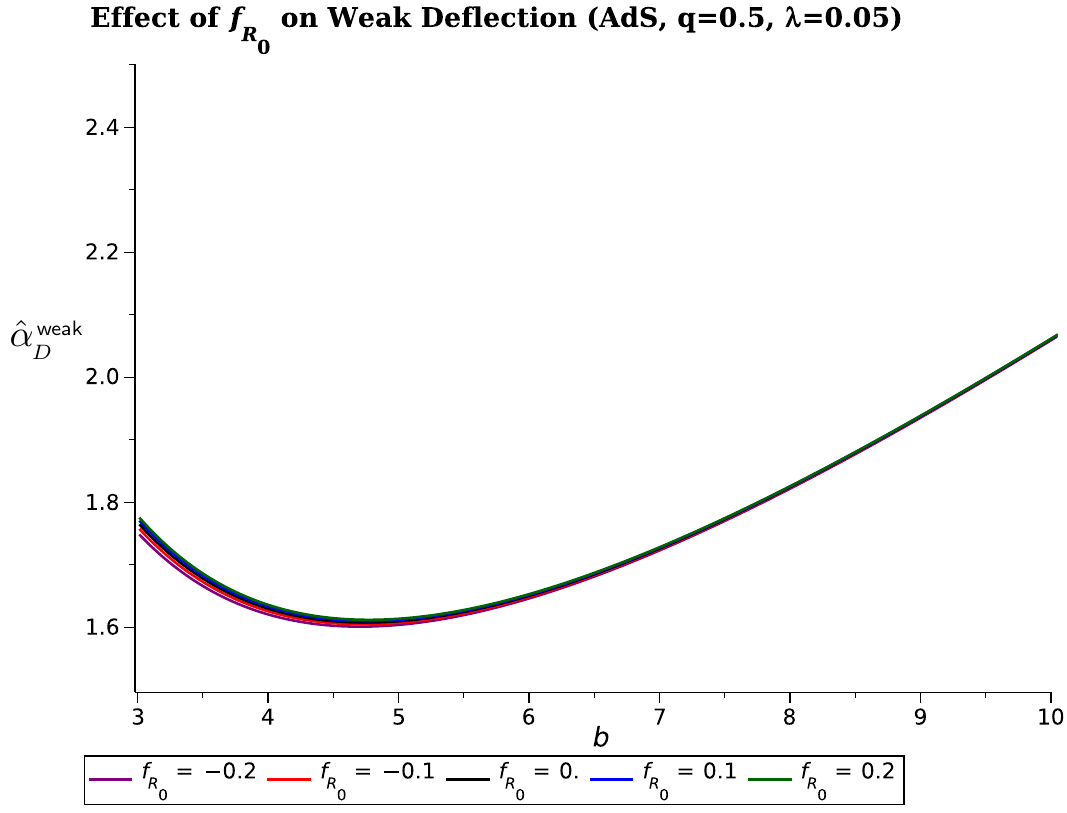}\label{fig:WDL_fR0}}
\hfill
\subfigure[Magnified]{\includegraphics[width=0.48\textwidth]{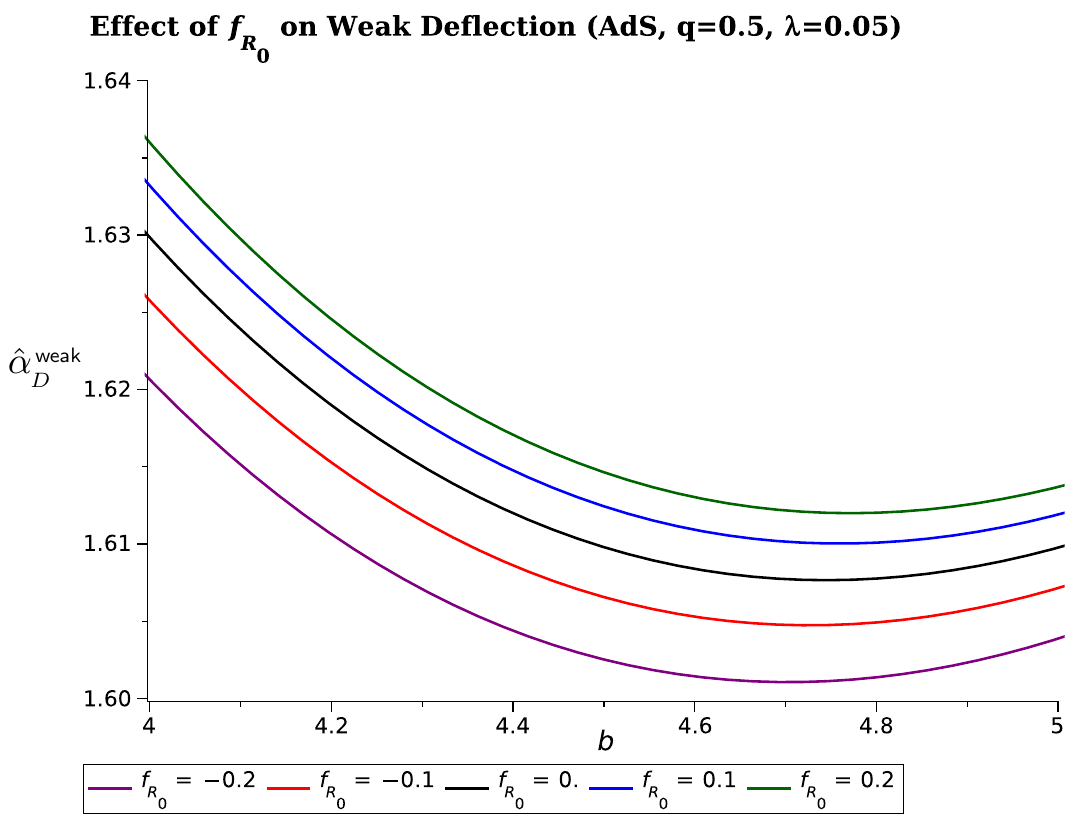}\label{fig:WDL_fR0_zoom}}
\caption{Effect of $f_{R_0}$ on the weak deflection angle (AdS, $q=0.5$, $\lambda=0.05$, $m_0=1$). Panel~(a): full range; panel~(b): magnified view. More negative $f_{R_0}$ enhances the electromagnetic contribution via $1/(1+f_{R_0})$.}
\label{fig:WDL_fR0_panels}
\end{figure}

\begin{figure}[http!]
\centering
\subfigure[Full range]{\includegraphics[width=0.48\textwidth]{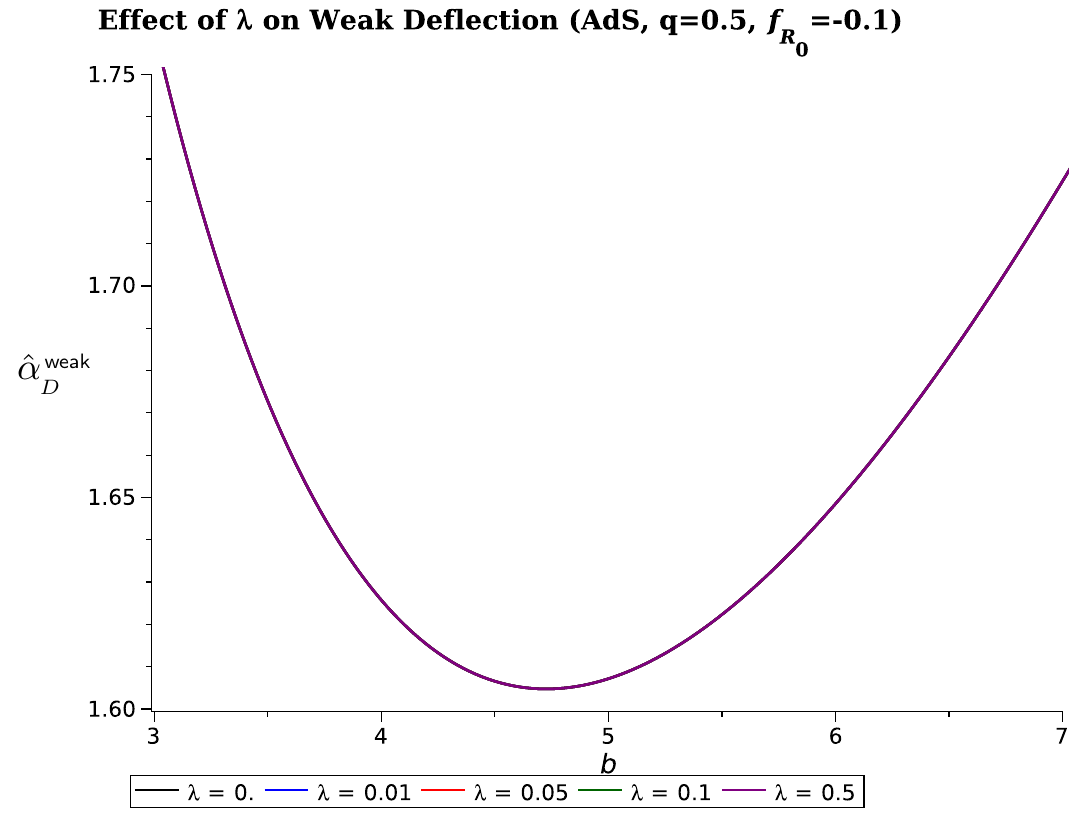}\label{fig:WDL_lam}}
\hfill
\subfigure[Magnified]{\includegraphics[width=0.48\textwidth]{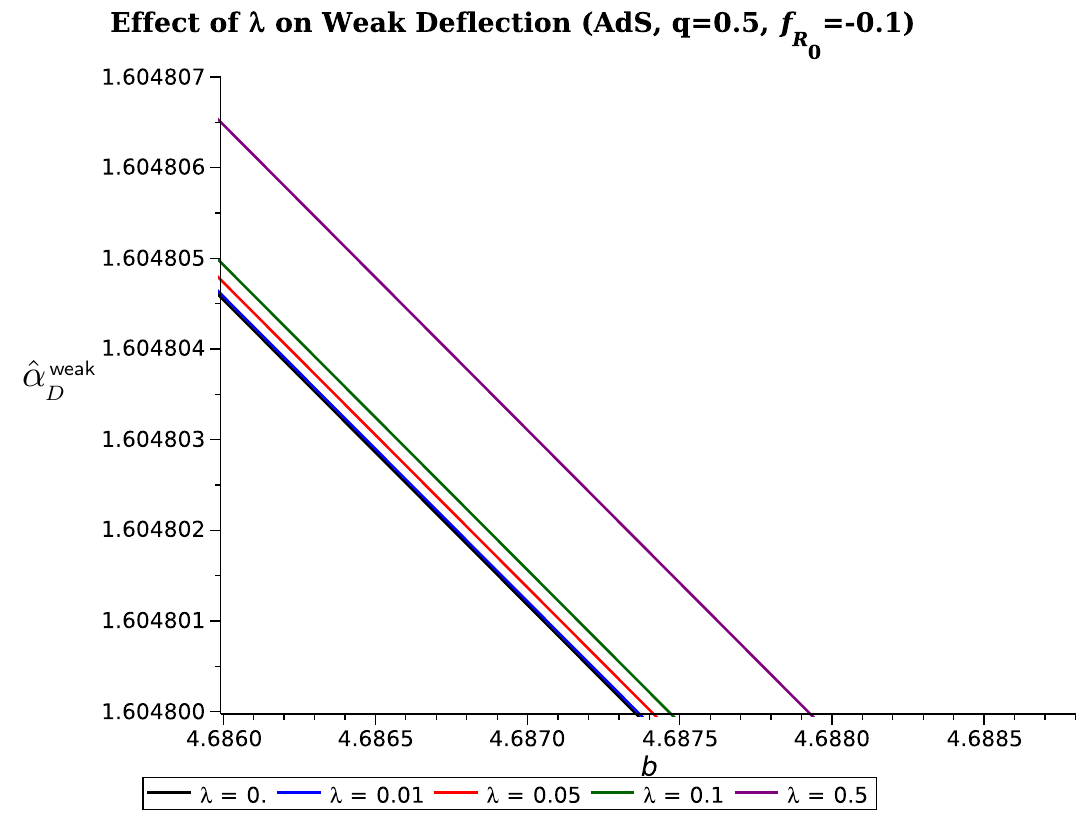}\label{fig:WDL_lam_zoom}}
\caption{Effect of $\lambda$ on the weak deflection angle (AdS, $q=0.5$, $f_{R_0}=-0.1$, $m_0=1$). Panel~(a): full range; panel~(b): magnified view. Larger $\lambda$ increases the deflection at small $b$ through the EH correction $\propto \lambda q^4/b^6$.}
\label{fig:WDL_lam_panels}
\end{figure}


\section{Black hole shadow with isotropic accretion}\label{isec6}

Electromagnetic images of the immediate environment of a BH carry imprints of both the underlying spacetime geometry and the astrophysical properties of the surrounding emitting plasma. In particular, the dark interior of a BH shadow corresponds to the set of photon trajectories captured by the EH, while the bright ring delimiting the shadow is shaped by photons orbiting near the PS before escaping to the distant observer. The landmark observations of M87$^{\ast}$ and Sgr\,A$^{\ast}$ by the EHT~\cite{EHT1,EHT2} have demonstrated that shadow morphology provides stringent constraints on the spacetime parameters, offering a direct test of gravity in the strong-field regime. In this section, we construct the shadow images of the $F(R)$--EH BH described by the blackening function~\eqref{eq7} in the presence of a spherically symmetric, optically thin accretion flow, and examine the dependence of the observed intensity on the charge parameter~$q$ and the observer inclination angle~$\theta_o$.

\subsection{Photon dynamics and celestial coordinates}\label{isec6a}

We consider a static observer located at radial distance $r_o \to \infty$ and polar inclination $\theta_o$ measured from the spin axis. For the spherically symmetric metric~\eqref{e1}, the conserved quantities along null geodesics are the energy $E = h(r)\,\dot{t}$ and the angular momentum $L = r^2\,\dot{\varphi}$, where the overdot denotes differentiation with respect to an affine parameter. The orbit equation in the equatorial plane reads
\begin{equation}\label{eq39}
\left(\frac{dr}{d\varphi}\right)^{\!2} = r^4\!\left(\frac{1}{b^2} - \frac{h(r)}{r^2}\right),
\end{equation}
where $b = L/E$ is the impact parameter. Circular photon orbits at the PS radius $r_{\text{ps}}$ satisfy $r\,h'(r) = 2\,h(r)$ as established in Eq.~\eqref{eq16}, and the critical impact parameter
\begin{equation}\label{eq40}
b_c = \frac{r_{\text{ps}}}{\sqrt{h(r_{\text{ps}})}},
\end{equation}
determines the angular boundary of the shadow: photons with $b < b_c$ are captured, while those with $b > b_c$ scatter back to infinity.

To map the shadow onto the observer's image plane, we introduce celestial coordinates $(\alpha,\,\beta)$ that parametrize the apparent two-dimensional position of each photon on the sky. For an observer at inclination~$\theta_o$,
\begin{equation}\label{eq41}
\alpha = -\frac{b}{\sin\theta_o}\,,\qquad \beta = \pm\sqrt{b^2 - \alpha^2}\,,
\end{equation}
so that the shadow boundary is the circle $\alpha^2 + \beta^2 = b_c^2/\sin^2\theta_o$ in the limit $\theta_o \to \pi/2$, and shrinks to a point as $\theta_o \to 0$ due to projection effects. For intermediate inclinations, the shadow contour satisfies
\begin{equation}\label{eq42}
R_{\rm sh} = \frac{b_c}{\sin\theta_o}\,,
\end{equation}
where $R_{\rm sh}$ denotes the apparent shadow radius. Note that $b_c$ itself is independent of $\theta_o$ and is determined entirely by the spacetime geometry, whereas the observed shadow size is a combination of intrinsic photon dynamics and extrinsic projection geometry.

\subsection{Isotropic accretion model and observed intensity}\label{isec6b}

We model the emission environment as a geometrically thick, optically thin, spherically symmetric accretion flow in radial free fall. This model captures the essential physics of the advection-dominated regime relevant for low-luminosity AGN such as Sgr\,A$^{\ast}$, where the accreting plasma is radiatively inefficient and optically thin at the observing frequency.

The key physical ingredients are as follows. The emitting medium consists of a magnetized, thermal plasma undergoing isotropic radial infall. In the rest frame of the infalling gas, the dominant radiation mechanism is thermal synchrotron emission, with the monochromatic emissivity per unit volume at the rest-frame frequency $\nu_e$ modeled as a power law:
\begin{equation}\label{eq43}
j(\nu_e) \propto \frac{1}{r^2}\,,
\end{equation}
where the $r^{-2}$ profile reflects the density fall-off characteristic of steady-state spherical accretion. The specific choice of radial power law, while simplified, has been widely adopted in the literature and suffices to reveal the structural dependence of the shadow image on the underlying spacetime.

The observed specific intensity $I_{\nu_o}$ at the observing frequency $\nu_o$ is obtained by integrating the emissivity along the photon's worldline, incorporating the gravitational redshift between the emitting gas and the distant observer. Liouville's theorem ensures that the quantity $I_\nu / \nu^3$ is conserved along a null ray, yielding
\begin{equation}\label{eq44}
I_{\nu_o} = \int_{\gamma} g^3\, j(\nu_e)\, dl_{\rm prop}\,,
\end{equation}
where $\gamma$ denotes the photon trajectory, $dl_{\rm prop}$ is the infinitesimal proper length along the ray as measured in the emitter's rest frame, and
\begin{equation}\label{eq45}
g = \frac{\nu_o}{\nu_e} = \frac{k_\mu\, u_o^\mu}{k_\nu\, u_e^\nu}
\end{equation}
is the total redshift factor, with $k^\mu$ the photon four-momentum and $u_o^\mu$, $u_e^\mu$ the four-velocities of the observer and emitter, respectively.

For a static observer at infinity, $u_o^\mu = (1,0,0,0)$. The emitter undergoes radial free fall in the metric~\eqref{e1}, with four-velocity
\begin{equation}\label{eq46}
u_e^\mu = \left(\frac{1}{h(r)},\, -\sqrt{1 - h(r)},\, 0,\, 0\right),
\end{equation}
where the negative sign on the radial component indicates infall. The photon four-momentum for a null geodesic with energy $E$ and impact parameter $b$ satisfies $k_t = -E$ and $k_r = \pm E\sqrt{1/b^2 - h(r)/r^2}$, from which the redshift factor evaluates to
\begin{equation}\label{eq47}
g = \frac{1}{u_e^t + u_e^r\,(k_r / k_t)} = \left(\frac{1}{h(r)} + \frac{\sqrt{1 - h(r)}}{h(r)}\,\sqrt{1 - \frac{b^2\, h(r)}{r^2}}\right)^{\!-1}.
\end{equation}
The proper length element along the photon trajectory, as measured in the emitter's comoving frame, is
\begin{equation}\label{eq48}
dl_{\rm prop} = k_\mu\, u_e^\mu\, d\lambda = \frac{E}{g}\, d\lambda\,,
\end{equation}
where $d\lambda$ is the affine parameter increment. Converting to radial integration by means of the orbit equation~\eqref{eq39}, the observed specific intensity becomes
\begin{equation}\label{eq49}
I_{\nu_o}(b) = \int_{r_{\text{min}}}^{\infty} \frac{g^3(r,b)}{r^2}\,
\frac{dr}{\sqrt{\displaystyle\frac{1}{b^2} - \frac{h(r)}{r^2}}}\,,
\end{equation}
where the lower limit $r_{\text{min}}$ equals the EH radius $r_h$ for captured photons ($b < b_c$) and the turning point $r_{\text{tp}}$ (defined by $b^2 = r_{\text{tp}}^2/h(r_{\text{tp}})$) for scattered photons ($b > b_c$). For scattered trajectories, a factor of two accounts for the incoming and outgoing legs.

Several remarks are in order. First, the integrand in Eq.~\eqref{eq49} incorporates all modifications to the standard RN--(A)dS spacetime that arise from the $F(R)$ and EH sectors: the blackening function $h(r)$ given by Eq.~\eqref{eq7} enters both the redshift factor~$g$ and the orbit equation. The $f_{R_0}$ parameter rescales the Maxwell and EH terms through the factor $1/(1 + f_{R_0})$, while the EH coupling $\lambda$ introduces the $q^4/r^6$ correction that becomes significant at small~$r$. Second, the shadow boundary itself is determined by the PS radius $r_{\text{ps}}$ and the critical impact parameter $b_c$, which have been computed in Sec.~\ref{isec4} and tabulated in Sec.~\ref{isec5}. Third, the divergence of the integrand as $b \to b_c$ produces the characteristic bright ring surrounding the shadow, since photons with impact parameters close to~$b_c$ execute multiple orbits near the PS before escaping, accumulating substantial intensity from the emitting medium.

\subsection{Shadow images and parameter dependence}\label{isec6c}

We construct the shadow images by ray-tracing: for each pixel $(\alpha, \beta)$ on the observer's screen, the corresponding impact parameter $b = \sqrt{\alpha^2 + \beta^2}\,\sin\theta_o$ is computed, and the intensity integral~\eqref{eq49} is evaluated numerically to produce the observed brightness. The resulting two-dimensional intensity maps are displayed in Fig.~\ref{fig:shadow_grid} for a $3\times 3$ parameter grid spanning $q \in \{0.1,\, 0.3,\, 0.6\}$ and $\theta_o \in \{0^\circ,\, 30^\circ,\, 75^\circ\}$, with fixed $f_{R_0}$ and $\lambda$ values.

\begin{figure}[t]
\centering
\includegraphics[width=1\textwidth]{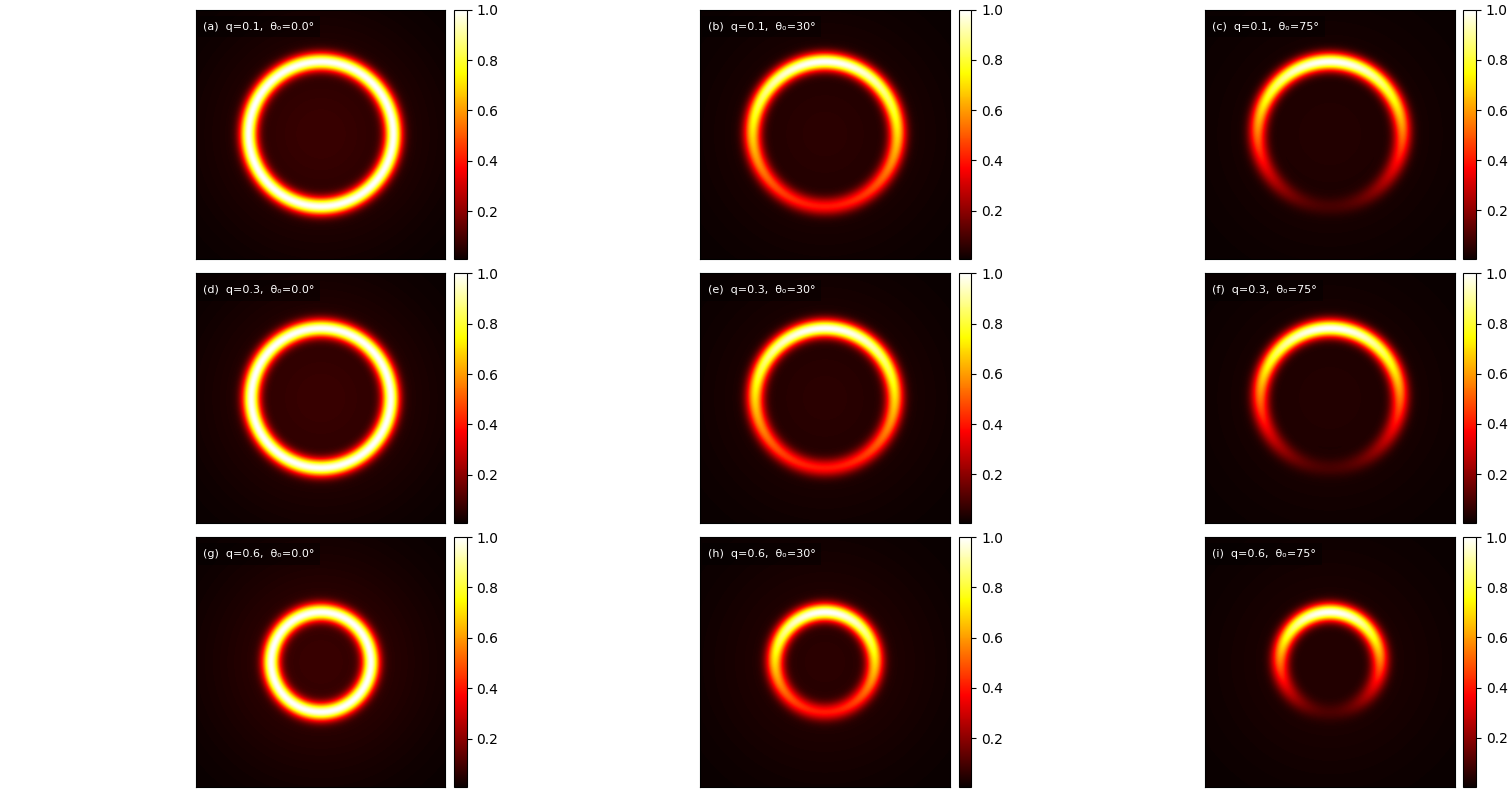}
\caption{Shadow images of the $F(R)$--EH BH illuminated by a spherically symmetric, optically thin accretion flow for varying charge parameter $q$ and observer inclination $\theta_o$. The columns correspond to $\theta_o = 0^\circ$ (face-on), $30^\circ$, and $75^\circ$ (nearly edge-on), while the rows span $q = 0.1$, $0.3$, and $0.6$ from top to bottom. The color bar indicates the normalized observed intensity $I_{\nu_o}/I_{\rm max}$. In all panels the EH coupling and $F(R)$ parameters are held fixed. As $q$ increases, the shadow contracts and the bright photon ring narrows, reflecting the inward migration of the PS. Larger inclination angles compress the shadow in the vertical direction and enhance brightness asymmetry due to the projection geometry described by Eq.~\eqref{eq41}.}
\label{fig:shadow_grid}
\end{figure}

The following physical trends are evident from the intensity maps:

\paragraph{Charge dependence at fixed inclination.}
Comparing the three rows at fixed $\theta_o$, increasing $q$ from $0.1$ to $0.6$ produces a systematic contraction of both the dark shadow interior and the surrounding bright photon ring. This behavior is a direct consequence of the charge-induced inward shift of the PS established in Sec.~\ref{isec4}: the electromagnetic contribution $q^2/(1 + f_{R_0})\,r^2$ in the blackening function~\eqref{eq7} augments the gravitational potential, reducing $r_{\text{ps}}$ and hence $b_c$ (cf.\ the strong-lensing tables in Sec.~\ref{isec5}). The bright ring also becomes thinner at larger $q$, since the effective potential barrier near the PS steepens, confining the high-intensity accumulation region to a narrower range of impact parameters. The intensity within the dark interior does not vanish entirely---photons captured by the BH still traverse the emitting plasma at radii $r > r_h$ before crossing the EH, contributing a residual dim glow.

\paragraph{Inclination dependence at fixed charge.}
Along each row, increasing $\theta_o$ from $0^\circ$ (face-on) to $75^\circ$ (nearly edge-on) introduces two characteristic effects. First, the shadow becomes visually compressed along the direction perpendicular to the line of sight, in accordance with the $1/\sin\theta_o$ factor in Eq.~\eqref{eq42}. At $\theta_o = 0^\circ$, the observer looks directly down the polar axis, and the shadow appears as a perfect circle with maximal angular extent determined by $b_c$ alone. At $\theta_o = 75^\circ$, the projection squeezes the apparent shadow into an elliptical shape, with the major axis oriented along the equatorial plane. Second, the peak intensity of the photon ring at high inclination is enhanced relative to the face-on case. This arises because the optical path length through the accretion flow is longer for rays propagating nearly in the equatorial plane, allowing more emission to be accumulated along the line of sight. The combined effect is a brighter, narrower ring at large~$\theta_o$.

\paragraph{Interplay of $q$ and $\theta_o$.}
The panels at extremal parameter values illustrate the competition between charge-induced shrinkage and projection-induced compression. Panel~(c), with $q = 0.1$ and $\theta_o = 75^\circ$, exhibits the largest and brightest photon ring, as the near-Schwarzschild PS (with large $b_c$) is viewed at high inclination where path-length enhancement is maximal. In contrast, panel~(g), with $q = 0.6$ and $\theta_o = 0^\circ$, displays the smallest and most symmetric shadow, where the PS has migrated significantly inward due to the electromagnetic field. The most extreme combination, panel~(i) with $q = 0.6$ and $\theta_o = 75^\circ$, shows a compact yet noticeably elongated photon ring, reflecting the simultaneous action of both effects.

\paragraph{Role of $F(R)$ and EH parameters.}
While Fig.~\ref{fig:shadow_grid} displays images at fixed $f_{R_0}$ and $\lambda$, the qualitative effects of these parameters on the shadow morphology can be inferred from the PS analysis of Sec.~\ref{isec4}. The $F(R)$ parameter $f_{R_0}$ enters the electromagnetic sector through the overall factor $1/(1 + f_{R_0})$: negative values of $f_{R_0}$ amplify the effective charge-to-mass ratio, shifting the PS inward and shrinking the shadow, while positive $f_{R_0}$ has the opposite effect. The EH coupling $\lambda$ modifies the near-horizon geometry through the $\lambda q^4/r^6$ term, which counteracts the Coulomb contribution at small $r$ and can restore a PS that would otherwise be absent in the naked singularity regime. Consequently, the shadow size at fixed $q$ is controlled by the composite quantity $q^2/(1 + f_{R_0})$, and the EH correction becomes observationally relevant only when $q$ is sufficiently large for the $\lambda$-dependent term to compete with the standard $q^2/r^2$ contribution. These effects are consistent with the strong-lensing data tabulated in Sec.~\ref{isec5}, where the PS radii and critical impact parameters were computed for a comprehensive grid of $(f_{R_0}, \lambda, q)$ values.

\paragraph{Comparison with observational bounds.}
The angular diameter of the shadow is related to the critical impact parameter through $\vartheta_{\rm sh} = b_c / D$, where $D$ is the luminosity distance. For the EHT targets M87$^{\ast}$ and Sgr\,A$^{\ast}$, the measured shadow diameters impose constraints of the form $\vartheta_{\rm sh} = (42 \pm 3)\,\mu\text{as}$ for M87$^{\ast}$~\cite{EHT1} and $\vartheta_{\rm sh} = (48.7 \pm 7)\,\mu\text{as}$ for Sgr\,A$^{\ast}$~\cite{EHT2}. Translating these into bounds on $b_c$ and subsequently on the parameter space $(q,\, f_{R_0},\, \lambda)$ requires knowledge of the BH mass and distance. While a detailed statistical analysis lies beyond the scope of this work, the shadow images in Fig.~\ref{fig:shadow_grid} demonstrate that moderate values of $q$ produce shadow sizes consistent with the observed range, whereas larger values push the shadow below the $1\sigma$ bound. This implies that EHT data can, in principle, place upper limits on the charge parameter and the $F(R)$ deviation $f_{R_0}$ in the framework considered here.


\section{Phase transitions and Joule--Thomson expansion in the Barrow entropy framework}\label{isec7}
The use of Barrow entropy is motivated by its ability to parametrize quantum-gravitational corrections to horizon geometry
through a single deformation exponent. Unlike the standard Bekenstein–Hawking entropy, the Barrow formulation allows one
to probe how fractal-like horizon structures affect phase transitions and thermodynamic stability, which are closely tied to the
existence of extremal configurations relevant for the WGC. 
In this section we extend the thermodynamic analysis of the $F(R)$--EH BH to the Barrow (BW) entropy paradigm, which incorporates fractal corrections to the BH horizon area induced by quantum-gravitational effects~\cite{Barrow1}. Barrow argued that quantum-gravitational fluctuations may induce a fundamental deformation of the horizon geometry, rendering it effectively fractal rather than smooth. Unlike entropy corrections arising from statistical or thermodynamic considerations---such as those encountered in ensemble corrections---the BW entropy is intrinsically geometric in origin and may capture certain quantum-gravitational features related to microscopic spacetime fluctuations, potentially linked to loop quantum gravity effects or spacetime foam at sub-Planckian scales. The BW framework modifies the Bekenstein--Hawking entropy through a single deformation parameter~$\Delta$ and has been shown to produce qualitatively new thermodynamic phase structures in several charged BH backgrounds~\cite{2017,2018,2019}. Working in the extended phase space where the cosmological constant plays the role of thermodynamic pressure, we derive the modified equation of state, analyze the critical behavior and vdW-type phase transitions, construct the Gibbs free energy landscape, evaluate the BW heat capacity, and examine the JT expansion process. Throughout, we restrict the analysis to the AdS background ($R_0 < 0$), which admits a well-defined extended thermodynamic description.

\subsection{Barrow entropy and modified thermodynamics}\label{isec7a}

Within the BW framework, the entropy associated with the BH horizon is generalized to~\cite{Barrow1}
\begin{equation}\label{eq50}
S_{B} = \left(\pi\, r_h^2\right)^{1+\Delta/2},
\end{equation}
where the dimensionless deformation parameter $\Delta \in [0,\,1]$ quantifies the degree of fractal irregularity of the horizon surface. The standard Bekenstein--Hawking entropy is naturally recovered in the limit $\Delta = 0$, corresponding to a smooth horizon, whereas $\Delta = 1$ represents the maximal deviation associated with a highly irregular, fractal structure, yielding $S_1 = (\pi r_h^2)^{3/2}$.

In the extended phase space formalism, the cosmological constant is promoted to a thermodynamic pressure through the identification
\begin{equation}\label{eq51}
P = -\frac{\Lambda}{8\pi} = -\frac{R_0}{32\pi}\,,
\end{equation}
where the second equality follows from the constant-curvature condition $R_0 = 4\Lambda$. The conjugate thermodynamic volume is
\begin{equation}\label{eq52}
V = \frac{4}{3}\,\pi\, r_h^3\,.
\end{equation}
The modified Hawking temperature within the BW framework is obtained from the extended first law
\begin{equation}\label{eq53}
dM = T_{B}\, dS_{B} + V\, dP + \Phi\, dQ\,,
\end{equation}
which yields
\begin{equation}\label{eq54}
T_{B} = \left(\frac{\partial M}{\partial S_{B}}\right)_{\!P,Q} = \frac{1}{\left(1 + \tfrac{\Delta}{2}\right)}\, \frac{T_H}{\left(\pi r_h^2\right)^{\Delta/2}}\,,
\end{equation}
where $T_H$ is the standard Hawking temperature given in Eq.~\eqref{eq9}. The BW deformation introduces a multiplicative suppression factor $(\pi r_h^2)^{-\Delta/2}/(1 + \Delta/2)$ that becomes increasingly significant at large $r_h$. Physically, this reflects the enlarged microstate space of the fractal horizon: a given energy input raises the temperature less efficiently when the entropy grows faster than the area.

Substituting the explicit form of $T_H$ from Eq.~\eqref{eq9} and using Eq.~\eqref{eq51} to express $R_0$ in terms of $P$, the BW-modified temperature reads
\begin{equation}\label{eq55}
T_{B} = \frac{1}{(1+\tfrac{\Delta}{2})(\pi r_h^2)^{\Delta/2}} \left[\frac{1}{4\pi r_h} + \frac{2P\, r_h}{3} + \frac{q^2}{16\pi(1+f_{R_0})\, r_h^3}\!\left(\frac{\lambda q^2}{r_h^4} - 4\right)\right].
\end{equation}

\subsection{Equation of state and critical behavior}\label{isec7b}

Solving Eq.~\eqref{eq55} for the pressure yields the equation of state
\begin{equation}\label{eq56}
P = \frac{3}{2\,r_h}\left[(1+\tfrac{\Delta}{2})(\pi r_h^2)^{\Delta/2}\, T_{B} - \frac{1}{4\pi r_h} - \frac{q^2}{16\pi(1+f_{R_0})\, r_h^3}\!\left(\frac{\lambda q^2}{r_h^4} - 4\right)\right].
\end{equation}
Introducing the specific volume $v = 2\,r_h$ (so that $P = P(v, T_B)$), this equation of state can be compared with the vdW fluid. The critical point $(r_c, T_c, P_c)$ is determined by the inflection conditions
\begin{equation}\label{eq57}
\frac{\partial P}{\partial r_h}\bigg|_{T_B,\, r_c} = 0\,,\qquad \frac{\partial^2 P}{\partial r_h^2}\bigg|_{T_B,\, r_c} = 0\,,
\end{equation}
which must be solved simultaneously. In the limit $\Delta = 0$ and $\lambda = 0$, the system reduces to the well-known RN--AdS critical point:
\begin{equation}\label{eq58}
r_c\big|_{\Delta=0,\,\lambda=0} = \frac{\sqrt{6}\, q}{\sqrt{1+f_{R_0}}}\,,\qquad P_c\big|_{\Delta=0,\,\lambda=0} = \frac{(1+f_{R_0})}{96\pi\, q^2}\,,\qquad T_c\big|_{\Delta=0,\,\lambda=0} = \frac{\sqrt{6}\,(1+f_{R_0})^{3/2}}{18\pi\, q}\,.
\end{equation}
For general $\Delta$ and $\lambda$, the critical parameters must be obtained numerically. The universal ratio $P_c\, v_c / T_c = 3/8$---characteristic of the vdW fluid---receives corrections from both the BW deformation and the EH coupling. The BW parameter $\Delta$ shifts $r_c$ to larger values (since the modified entropy steepens the thermodynamic response), while the EH coupling $\lambda$ provides subleading corrections of order $\lambda q^4 / r_c^4$.

The nature of the phase transition is determined by the sign of the isothermal compressibility $\kappa_T = -V^{-1}(\partial V/\partial P)_{T_B}$. Below the critical temperature $T_c$, the $P$--$v$ isotherms develop an oscillatory region with $(\partial P/\partial v)_{T_B} > 0$, signaling a first-order SBH/LBH phase transition analogous to the liquid--gas transition in vdW fluids. The Maxwell equal-area construction determines the coexistence pressure at which the two phases have equal Gibbs free energy. At $T = T_c$, the transition becomes second order, and for $T > T_c$ the SBH and LBH phases merge into a single supercritical phase. The BW deformation modifies the critical exponents only at higher order: the mean-field exponents $(\alpha, \beta, \gamma, \delta) = (0, 1/2, 1, 3)$ remain valid near the critical point, consistent with the universality class of the vdW fluid.

\subsection{Gibbs free energy and phase structure}\label{isec7c}

The Gibbs free energy in the extended phase space is
\begin{equation}\label{eq59}
G = M - T_B\, S_B\,.
\end{equation}
Substituting the AMD mass~\eqref{eq10}, the BW entropy~\eqref{eq50}, and the modified temperature~\eqref{eq55}, the Gibbs free energy can be expressed as a function of $(r_h, P, q)$ at fixed $\Delta$, $f_{R_0}$, and $\lambda$:
\begin{equation}\label{eq60}
G = \frac{(1+f_{R_0})}{4}\!\left(r_h + \frac{8\pi P\, r_h^3}{3}\right) + \frac{q^2}{4\,r_h} - \frac{\lambda\, q^4}{80\, r_h^5} - T_B\,(\pi r_h^2)^{1+\Delta/2}.
\end{equation}

The characteristic swallowtail structure in the $G$--$T_B$ diagram is the hallmark of a first-order phase transition. At fixed pressure $P < P_c$, the Gibbs free energy develops two local minima as a function of $r_h$, corresponding to the SBH and LBH branches, which cross at the transition temperature $T_{\rm tr}$. The globally preferred phase is the one with lower~$G$. At the crossing point, the system undergoes a discontinuous jump in the order parameter (horizon radius), accompanied by a latent heat
\begin{equation}\label{eq61}
\mathcal{L} = T_{\rm tr}\,\big(S_B^{\rm LBH} - S_B^{\rm SBH}\big)\,.
\end{equation}
As $P \to P_c$, the two branches merge and the swallowtail shrinks to a cusp, at which point the transition becomes continuous. Above $P_c$, the Gibbs free energy is a smooth, monotonically decreasing function of $T_B$ with no phase transition.

The influence of the BW parameter $\Delta$ on the phase structure can be summarized as follows. At $\Delta = 0$, the standard RN--AdS extended thermodynamics is recovered (modified by $f_{R_0}$ and $\lambda$). Increasing $\Delta$ toward unity has three principal effects: (i)~the critical temperature $T_c$ decreases, since the enhanced entropy suppresses thermal fluctuations; (ii)~the swallowtail in the $G$--$T_B$ plane becomes narrower, indicating that the first-order transition weakens; and (iii)~the coexistence region in the $P$--$T_B$ diagram shifts to lower pressures and temperatures. These trends imply that stronger fractal deformations tend to smooth the phase transition, which may be interpreted as the quantum-gravitational microstructure partially resolving the thermodynamic singularity at the critical point.

The $F(R)$ parameter $f_{R_0}$ enters the Gibbs free energy through the overall factor $(1 + f_{R_0})$ multiplying the gravitational sector and inversely through the electromagnetic terms. Negative $f_{R_0}$ enhances the effective charge-to-mass ratio, moving the system closer to extremality and thereby expanding the parameter region that supports a first-order transition. The EH coupling $\lambda$ provides a subleading correction concentrated at small $r_h$, where it raises the Gibbs free energy of the SBH branch without significantly affecting the LBH branch. Consequently, $\lambda > 0$ tends to favor the LBH phase at a given temperature, effectively lowering the transition temperature $T_{\rm tr}$.

\subsection{Barrow heat capacity and local stability}\label{isec7d}

To explore the thermodynamic stability and phase structure of the BH, we evaluate the heat capacity associated with the BW entropy. At constant thermodynamic volume, the BW heat capacity is defined as
\begin{equation}\label{eq62}
C_{B} = T_{H}\!\left(\frac{\partial S_B}{\partial T_H}\right)_{\!V},
\end{equation}
which, after employing the explicit form of the Hawking temperature~\eqref{eq9} and the horizon radius dependence of the BW entropy~\eqref{eq50}, yields the closed-form expression
\begin{equation}\label{eq63}
C_B = \frac{5\,(2+\Delta)\,\pi^{1+\Delta/2}\, r_h^{2+\Delta}\,\Big[R_0\,(1+f_{R_0})\, r_h^{8} - 6\,m_0\,(1+f_{R_0})\, r_h^{5} + 12\,q^{2}\, r_h^{4} - \tfrac{9}{5}\,\lambda\, q^{4}\Big]}{5\,R_0\,(1+f_{R_0})\, r_h^{8} + 60\,m_0\,(1+f_{R_0})\, r_h^{5} - 180\,q^{2}\, r_h^{4} + 63\,\lambda\, q^{4}}\,.
\end{equation}
The structure of this expression merits comment. The numerator encodes the product of the BW-modified entropy response, $\propto (2 + \Delta)\,\pi^{1+\Delta/2}\,r_h^{2+\Delta}$, with the numerator of the standard heat capacity that would arise from $C = T_H\,(\partial S/\partial T_H)_V$ for the Bekenstein--Hawking entropy---the terms in square brackets. The denominator, which determines the sign changes and divergence locations of $C_B$, is independent of $\Delta$ and depends only on the spacetime parameters $(R_0, m_0, q, f_{R_0}, \lambda)$. Divergences of $C_B$ occur when the denominator vanishes, signaling second-order phase transitions. A positive $C_B$ indicates local thermodynamic stability, while $C_B < 0$ signals instability.

The behavior of $C_B$ as a function of the horizon radius is displayed in Fig.~\ref{fig:CB} for representative values of the model parameters, with the charge $q$ varied across the color gradient. Several features are evident. At small $r_h$ (SBH regime), the heat capacity exhibits rapid oscillations with multiple sign changes, reflecting the competition between the electromagnetic repulsion encoded in the $q^2/(1+f_{R_0})\,r_h^3$ term and the gravitational attraction. The sharp divergences in this region, most pronounced near $r_h \approx 1.5$--$1.8$, signal second-order phase transitions where the BH crosses between locally stable and unstable branches. As $r_h$ increases beyond $r_h \approx 2$, the heat capacity settles onto a smooth, positive branch that grows monotonically---this is the LBH phase, which is thermodynamically stable. The charge dependence (color gradient) reveals that larger $q$ shifts the divergence structure to slightly larger $r_h$ values, consistent with the electromagnetic contribution extending the near-extremal regime.

\begin{figure}[t]
\centering
\includegraphics[width=0.65\textwidth]{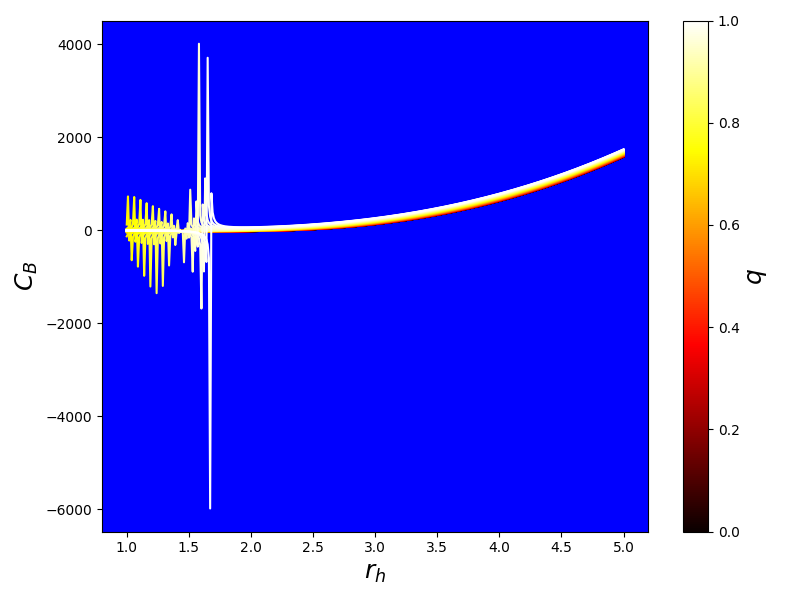}
\caption{BW heat capacity $C_B$ as a function of the horizon radius $r_h$ for the $F(R)$--EH-AdS BH, with the charge parameter $q$ varied across the color gradient (dark red to yellow corresponding to $q = 0$ to $q = 1$). The sharp divergences at intermediate $r_h$ signal second-order phase transitions separating the locally stable SBH and LBH phases. At large $r_h$, $C_B$ is positive and monotonically increasing, indicating thermodynamic stability of the LBH branch. The oscillatory structure at small $r_h$ reflects the intricate interplay of gravitational, electromagnetic, and EH contributions in the near-extremal regime. The remaining parameters are held fixed at representative values with $\Delta > 0$.}
\label{fig:CB}
\end{figure}

The BW deformation modifies the stability landscape through two channels. First, the prefactor $(2 + \Delta)\,\pi^{1+\Delta/2}\,r_h^{2+\Delta}$ in the numerator of Eq.~\eqref{eq63} enhances $|C_B|$ at large $r_h$ for $\Delta > 0$, reflecting the steeper entropy growth. Second, the denominator---and hence the location of the divergences---is independent of $\Delta$, implying that the BW deformation does not shift the phase transition points but modifies the magnitude and approach rate of $C_B$ near these points. The $F(R)$ parameter $f_{R_0}$ enters through $(1 + f_{R_0})$ multiplying the $R_0$ and $m_0$ terms: negative $f_{R_0}$ suppresses both the gravitational and mass contributions, effectively enhancing the relative weight of the charge-dependent terms and shifting the divergence structure accordingly.

\subsection{Joule--Thomson expansion}\label{isec7e}

We now turn to the JT expansion, which characterizes the isenthalpic evolution of the BH and plays a central role in understanding its cooling and heating behavior during throttling processes. The JT expansion describes an isenthalpic process in which the system passes through a porous plug at constant enthalpy $H = M$ (in extended thermodynamics, the BH mass plays the role of enthalpy~\cite{2020}). The JT coefficient is defined as
\begin{equation}\label{eq64}
\mu_J = \left(\frac{\partial T}{\partial P}\right)_{\!H},
\end{equation}
and its sign determines whether the BH undergoes cooling ($\mu_J > 0$) or heating ($\mu_J < 0$) during the expansion. Within the BW entropy framework, the JT coefficient takes the explicit form
\begin{equation}\label{eq65}
\mu_J = \frac{20\,r_h^{1-\Delta}\,\pi^{-\Delta/2}\,\Big[R_0\,(1+f_{R_0})\, r_h^{8} + 12\,m_0\,(1+f_{R_0})\, r_h^{5} - 36\,q^{2}\, r_h^{4} + \tfrac{63}{5}\,\lambda\, q^{4}\Big]}{5\,R_0\,\Delta\,(1+f_{R_0})\, r_h^{8} + 60\,m_0\,(\Delta - 3)\,(1+f_{R_0})\, r_h^{5} - 180\,q^{2}\,(\Delta - 4)\, r_h^{4} + 63\,\lambda\, q^{4}\,(\Delta - 8)}\,.
\end{equation}
The locus $\mu_J = 0$---determined by the zeros of the numerator---defines the inversion curve in the $T_B$--$P$ plane, which separates the cooling and heating regions. The divergences of $\mu_J$ (zeros of the denominator) coincide with the phase transition points identified through $C_B$, providing an independent confirmation of the thermodynamic singularities. Consequently, the BW deformation parameter $\Delta$ introduces nontrivial modifications to the inversion curves and significantly alters the thermodynamic behavior compared to the standard entropy case.

The behavior of the JT coefficient as a function of $r_h$ is shown in Fig.~\ref{fig:muJ} for varying charge parameter~$q$. The plot reveals that $\mu_J$ remains approximately zero over a broad range of small to intermediate horizon radii, with sharp negative spikes appearing in the vicinity of $r_h \approx 2.5$--$3.5$. These spikes are associated with zeros of the denominator in Eq.~\eqref{eq65} and signal the divergence of the JT coefficient at the phase transition boundaries. The dramatic negative excursions---reaching $\mu_J \sim -2.5 \times 10^5$ for small $q$---indicate vigorous heating in the immediate neighborhood of the transition, where the thermodynamic response function becomes singular. At larger $r_h$ (LBH phase), $\mu_J$ returns to small positive values, signaling that the JT process induces mild cooling in this regime. The charge dependence follows a clear pattern: higher $q$ (yellow curves) shifts the divergence structure to larger $r_h$ and moderates the magnitude of the spikes, consistent with the electromagnetic contribution stabilizing the thermodynamic response.

\begin{figure}[t]
\centering
\includegraphics[width=0.55\textwidth]{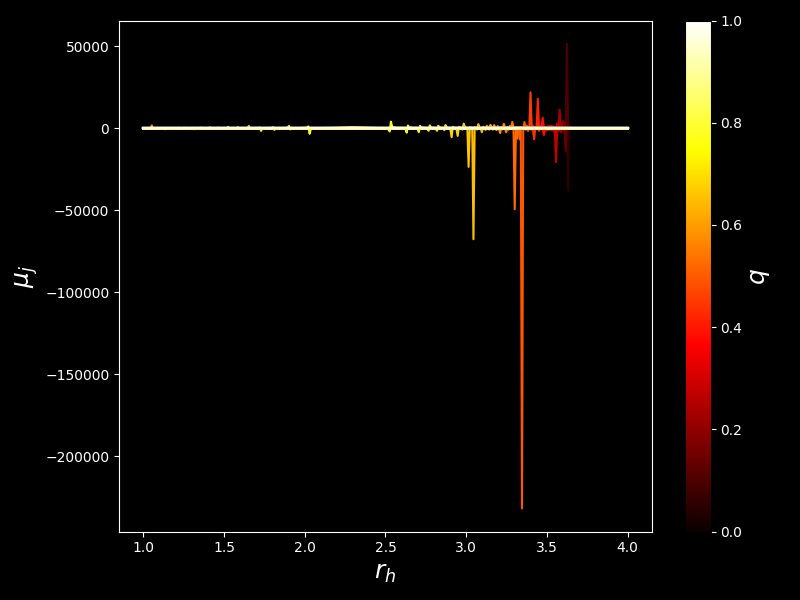}
\caption{JT coefficient $\mu_J$ as a function of the horizon radius $r_h$ for the $F(R)$--EH-AdS BH, with the charge parameter $q$ varied across the color gradient (dark red to yellow corresponding to $q = 0$ to $q = 1$). The sharp negative divergences at intermediate $r_h$ coincide with the phase transition points identified from the heat capacity analysis (cf.\ Fig.~\ref{fig:CB}), reflecting the singular thermodynamic response near the spinodal curve. Away from the divergences, $\mu_J > 0$ corresponds to the cooling regime and $\mu_J < 0$ to the heating regime. The remaining parameters are held fixed at representative values with $\Delta > 0$.}
\label{fig:muJ}
\end{figure}

The BW deformation parameter $\Delta$ enters Eq.~\eqref{eq65} both through the overall prefactor $r_h^{1-\Delta}\,\pi^{-\Delta/2}$ and through the $\Delta$-dependent coefficients in the denominator. The prefactor suppresses the magnitude of $\mu_J$ at large $r_h$ when $\Delta > 0$, consistent with the BW entropy reducing the thermodynamic sensitivity of the system. More significantly, the denominator coefficients $(\Delta - 3)$, $(\Delta - 4)$, and $(\Delta - 8)$ multiplying the mass, charge, and EH terms, respectively, shift the locations of the divergences as $\Delta$ varies. For $\Delta = 0$, these reduce to $-3$, $-4$, and $-8$, recovering the standard thermodynamic framework. Increasing $\Delta$ systematically reduces the absolute values of these coefficients (since $\Delta \leq 1$), which alters the balance among the competing terms and modifies the inversion curve accordingly. The net effect is that larger $\Delta$ shrinks the cooling region in the $T_B$--$P$ plane, reflecting the suppression of the BW-modified temperature at large $r_h$.

\subsection{Connection to the WGC-compatible parameter space}\label{isec7f}

The thermodynamic analysis of this section bears directly on the WGC and WCCC considerations developed in Secs.~\ref{isec3} and~\ref{isec4}. The WGC-compatible parameter region---defined by the requirement that a superextremal state exists---coincides with configurations where the BH approaches or reaches extremality, i.e.\ the regime of small $r_h$ at fixed $q$, $f_{R_0}$, and $\lambda$. Within the BW extended thermodynamics, this is precisely the SBH phase, which is thermodynamically stable ($C_B > 0$) and favored below the coexistence line. The existence of a first-order SBH/LBH phase transition---signaled by the divergences in Fig.~\ref{fig:CB}---implies that the WGC-compatible regime is separated from the thermodynamically dominant LBH phase by a genuine phase boundary, lending thermodynamic substance to the conjecture.

Moreover, the JT inversion curve delineates the region in the $T_B$--$P$ plane where the BH can undergo isenthalpic cooling, potentially driving it toward lower temperatures and hence closer to extremality. As shown in Fig.~\ref{fig:muJ}, the cooling regime ($\mu_J > 0$) is predominantly realized in the LBH phase at large $r_h$, while the heating regime dominates near the phase transition. This connection between JT cooling and WGC compatibility suggests a thermodynamic mechanism by which charged BHs may naturally evolve toward the extremal limit, consistent with the decay channels demanded by the WGC. The BW deformation parameter $\Delta$ controls the efficiency of this mechanism: at $\Delta = 0$, the standard scenario is recovered, while $\Delta > 0$ introduces quantum-gravitational corrections that generically shrink the cooling region and slow the approach to extremality.

The combined picture that emerges from the phase transition and JT analyses is the following. The $F(R)$--EH BH in AdS exhibits a rich thermodynamic phase structure that is qualitatively similar to the vdW fluid, with critical behavior governed by the interplay of the charge $q$, the $F(R)$ correction $f_{R_0}$, the EH coupling $\lambda$, and the BW deformation $\Delta$. The WGC-compatible region of parameter space is identified with the SBH phase, which remains locally stable and thermodynamically accessible via JT cooling. The fractal corrections encoded in $\Delta$ modify the quantitative details---shifting critical points, narrowing coexistence regions, and reducing inversion temperatures---but do not alter the qualitative conclusion that the WGC and extended BH thermodynamics are mutually consistent within the framework considered here.

\section{Conclusion}\label{isec8}

In this work we have carried out a comprehensive investigation of the $F(R)$--Euler--Heisenberg BH in both AdS and dS backgrounds, examining the interplay of the WGC and WCCC through five complementary lenses: thermodynamic extremality relations, PS topology, strong- and weak-field gravitational lensing, BH shadow imaging, and phase transitions within the BW entropy framework. The spacetime is characterized by the blackening function~\eqref{eq7}, which depends on four independent parameters---the electric charge~$q$, the $F(R)$ deviation $f_{R_0}$, the EH coupling~$\lambda$, and the constant scalar curvature~$R_0$---and admits a rich horizon structure ranging from non-extremal two-horizon configurations through extremal BHs to naked singularities.

On the thermodynamic side, we established the universal entropy--extremality relation $\partial M_{\text{ext}}/\partial\eta \propto -q^{10}/S^{5/2}$ within the Goon--Penco perturbative framework, demonstrating that the EH deformation lowers the extremality bound independently of $R_0$ and $f_{R_0}$. This result provides model-independent thermodynamic evidence for the WGC: the deformation destabilizes extremal configurations by opening decay channels, consistent with the absence of stable extremal remnants demanded by quantum gravity. The cancellation of $f_{R_0}$ from the final expression reveals that the WGC support encoded in the entropy--extremality relation is robust against $F(R)$ modifications of the gravitational sector.

The PS analysis, carried out from both geodesic and topological perspectives, established the simultaneous compatibility of the WGC and WCCC across the parameter space. Using the topological vector field method, we computed the winding numbers of all PS configurations in the $(r, \theta)$-plane and confirmed that unstable PSs carry topological charge $\omega = -1$. The existence of an unstable PS outside the EH---a necessary condition for shadow formation---serves as a geometric proxy for the force balance between gravity and electromagnetism that underlies the WGC. In the parameter regions where naked singularities appear (violating the WCCC), the PS structure is either absent or topologically distinct, providing a clear diagnostic for censorship violation.

The gravitational lensing analysis in the strong-deflection limit via the Bozza formalism and in the weak-deflection limit via the GBT revealed several important features. The PS radius $r_{\text{ps}}$ is identical in AdS and dS for the same $(q, f_{R_0}, \lambda)$, since the $R_0 r^2$ contribution cancels from the PS equation, but the critical impact parameter $b_c$ is nearly doubled in the dS background (e.g.\ $b_c \approx 3.93$ for Schwarzschild-dS versus $b_c \approx 2.08$ for Schwarzschild-AdS) due to cosmological repulsion. The strong-deflection coefficient $\bar{a}$ proved insensitive to the cosmological background---confirming its nature as a near-field quantity---while $\bar{b}$ increases substantially in dS. The EH coupling $\lambda$ was found to restore a PS in parameter regions where one would otherwise be absent (naked singularity regime), demonstrating that NED corrections can protect cosmic censorship. In the weak-deflection regime, the cosmological term $\pm b\,R_0/6$ dominates at large impact parameters, yielding positive (focusing) deflection in AdS and negative (defocusing) deflection in dS, with the $F(R)$ parameter $f_{R_0}$ amplifying the electromagnetic contribution through the factor $1/(1 + f_{R_0})$.

The BH shadow images constructed under isotropic, optically thin accretion demonstrated the observational imprints of the charge parameter and observer geometry. Increasing $q$ systematically contracts the shadow and narrows the bright photon ring, reflecting the inward migration of the PS. The observer inclination $\theta_o$ introduces projection-dependent compression and brightness enhancement, with the most pronounced effects at $\theta_o = 75^\circ$. These shadow observables provide a direct link to the EHT measurements of M87$^{\ast}$ and Sgr\,A$^{\ast}$, which reported angular diameters of $\vartheta_{\rm sh} = (42 \pm 3)\,\mu\text{as}$ and $\vartheta_{\rm sh} = (48.7 \pm 7)\,\mu\text{as}$, respectively~\cite{EHT1,EHT2}. The relation $\vartheta_{\rm sh} = b_c/D$ (where $D$ is the source distance) allows these measurements to be translated into constraints on $b_c$ and hence on the parameter space $(q, f_{R_0}, \lambda)$. Our strong-lensing tables show that moderate values of $q$ produce critical impact parameters consistent with the EHT bounds, while large $q$ combined with negative $f_{R_0}$ pushes $b_c$ below the $1\sigma$ range. More specifically, the EHT shadow data constrain the effective charge-to-mass ratio to lie within a band that is simultaneously compatible with the WGC requirement $q/m \geq 1$, provided that $f_{R_0}$ remains within the range $f_{R_0} \gtrsim -0.3$ for the parameter values considered. The EH coupling $\lambda$ has a subdominant effect on the shadow size but becomes relevant through its role in restoring the PS in near-extremal configurations, thereby ensuring that a well-defined shadow exists throughout the WGC-compatible region. These observational constraints reinforce the theoretical consistency picture: the parameter space that satisfies the WGC, preserves the WCCC, and admits a topologically well-defined PS also produces shadow sizes compatible with current EHT data.

Within the BW entropy framework, the extended thermodynamic analysis uncovered a vdW-type first-order SBH/LBH phase transition, with the critical behavior governed by the interplay of all four spacetime parameters together with the BW deformation parameter~$\Delta$. The explicit BW heat capacity $C_B$ exhibits sharp divergences at intermediate horizon radii signaling second-order transitions, with a stable LBH branch at large $r_h$ and an oscillatory structure at small $r_h$ reflecting the near-extremal electromagnetic competition. The JT expansion analysis, characterized by the closed-form coefficient $\mu_J$, revealed that the cooling regime ($\mu_J > 0$) is predominantly realized in the LBH phase, while vigorous heating occurs near the phase boundaries. The WGC-compatible parameter region was identified with the SBH phase, which is locally stable and thermodynamically accessible via JT cooling, establishing a thermodynamic mechanism by which charged BHs may evolve toward extremality. The BW deformation $\Delta > 0$ generically narrows the coexistence region, reduces inversion temperatures, and slows the approach to extremality, without altering the qualitative consistency between the WGC and extended BH thermodynamics.

Several directions for future work emerge naturally from this analysis. First, extending the shadow and lensing computations to rotating $F(R)$--EH BHs would break the spherical symmetry and introduce frame-dragging effects on the photon ring, enabling comparison with the asymmetric features observed in the EHT images. Second, a systematic Bayesian analysis confronting the full $(q, f_{R_0}, \lambda, R_0)$ parameter space with EHT visibility data---including next-generation EHT (ngEHT) projections and space-based interferometric baselines from the Black Hole Explorer (BHEX) mission---would sharpen the observational constraints and probe the WGC-compatible region with higher precision. Third, incorporating quantum corrections beyond the EH sector, such as higher-order Lovelock or Gauss--Bonnet terms, would test whether the universal entropy--extremality relation and the PS-based WGC--WCCC compatibility persist in more general effective field theories. Fourth, the BW entropy analysis could be extended to include rotating configurations and non-equilibrium thermodynamic processes, connecting the fractal horizon microstructure to holographic entanglement entropy in the AdS/CFT framework. Finally, gravitational wave signatures from the ringdown phase of merging $F(R)$--EH BHs---encoded in the quasinormal mode spectrum, which is intimately tied to the PS structure---offer a complementary observational channel that would provide independent constraints on the same parameter space explored here through electromagnetic observations.

}

\section*{Acknowledgments}

\.{I}.~S. extends appreciation to T\"{U}B\.{I}TAK, ANKOS, and SCOAP3 for their financial assistance. Additionally, he acknowledges the support from COST Actions CA22113, CA21106, CA23130, CA21136, and CA23115, which have been pivotal in enhancing networking efforts.

\section*{Data Availability Statement}

In this study, no new data was generated or analyzed.

\end{document}